\documentclass[final,5p,times]{elsarticle}
\usepackage{lineno,hyperref}
\modulolinenumbers[5]
\usepackage{amsmath,amssymb,amsfonts}
\usepackage{algorithmic}
\usepackage{graphicx}
\DeclareGraphicsExtensions{.pdf,.jpeg,.png}
\usepackage{subfigure}
\usepackage{epsfig}
\usepackage{booktabs} 
\usepackage{multirow}
\usepackage{array}
\usepackage{dirtree}
\usepackage{epstopdf}
\usepackage{caption,setspace}
\usepackage{wrapfig}

\usepackage{graphics}
\usepackage{diagbox}

\journal{Journal of \LaTeX\ Templates}

\biboptions{sort&compress}
\begin{document}

\begin{frontmatter}

  \title{Reversible data hiding in encrypted images based on pixel prediction and multi-MSB planes rearrangement}

  \author[1]{Zhaoxia Yin}
  \author[1]{Xiaomeng She}
  \address[1]{Anhui Province Key Laboratory of Multimodal Cognitive Computation, School of Computer Science and Technology, Anhui University, 230601, P.R.China}
  \author[1]{Jin Tang\corref{mycorrespondingauthor}}
  \ead{tangjin@ahu.edu.cn}
  \cortext[mycorrespondingauthor]{Corresponding author}
  \author[1]{Bin Luo}
  \begin{abstract}
    Great concern has arisen in the field of reversible data hiding in encrypted images (RDHEI) due to the development of cloud storage and privacy protection. RDHEI is an effective technology that can embed additional data after image encryption, extract additional data error-free and reconstruct original images losslessly. In this paper, a high-capacity and fully reversible RDHEI method is proposed, which is based on pixel prediction and multi-MSB (most significant bit) planes rearrangement. First, the median edge detector (MED) predictor is used to calculate the predicted value. Next, unlike previous methods, in our proposed method, signs of prediction errors (PEs) are represented by one bit plane and absolute values of PEs are represented by other bit planes. Then, we divide bit planes into uniform blocks and non-uniform blocks, and rearrange these blocks. Finally, according to different pixel prediction schemes, different numbers of additional data are embedded adaptively. The experimental results prove that our method has higher embedding capacity compared with state-of-the-art RDHEI methods.
  \end{abstract}


\end{frontmatter}

\section{Introduction}
\label{sec::introduction}
\par Data hiding is a technique that can embed data into multimedia and extract the data error-free by modifying the cover medium slightly. In order to protect the cover medium, reversible data hiding (RDH) is proposed~\cite{barton1997method}, which can reconstruct the cover medium losslessly. In the past several decades, RDH has attracted increasing attention, and has been applied in many fields, such as military communication, medical diagnosis, judicial evidence obtaining and so on~\cite{shi2016reversible}.
\par A major current focus in RDH is how to ensure low distortion of original images after embedding data, therefore, visual quality of marked images is the key metric of RDH methods. To achieve better visual quality, many RDH methods have been proposed in the past several decades~\cite{celik2005lossless,celik2006lossless,zhang2013recursive,ni2006reversible,li2013general,jia2019reversible,ou2019high,gao2020reversible,tian2003reversible,alattar2004reversible,kim2008novel}. These methods are mainly divided into three categories: lossless compression~\cite{celik2005lossless,celik2006lossless,zhang2013recursive}, histogram shifting~\cite{ni2006reversible,li2013general,jia2019reversible,ou2019high,gao2020reversible} and difference expansion~\cite{tian2003reversible,alattar2004reversible,kim2008novel}. The first category, lossless compression, is used in many RDH methods to vacate room for data embedding. To achieve better performance, lots of RDH methods based on histogram shifting have been proposed. The central idea of these methods is to use the peak points and minimum points of the histogram of an image, and embed additional data by modifying the grayscale values. The third category is based on difference expansion, which embeds data by expanding the difference between two pixels.
\par
\begin{figure*}[!ht]
  \centering
  \subfigure[]
  {
    \label{fig-1-a}
    \includegraphics[width=1\textwidth]{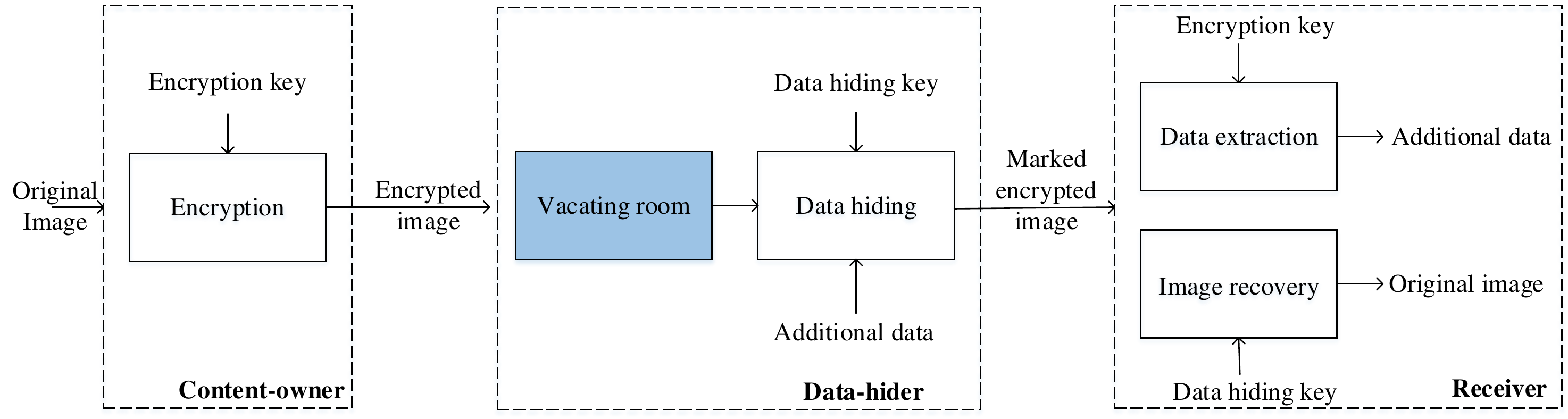}
  }
  \subfigure[]
  {
    \label{fig-1-b}
    \includegraphics[width=1\textwidth]{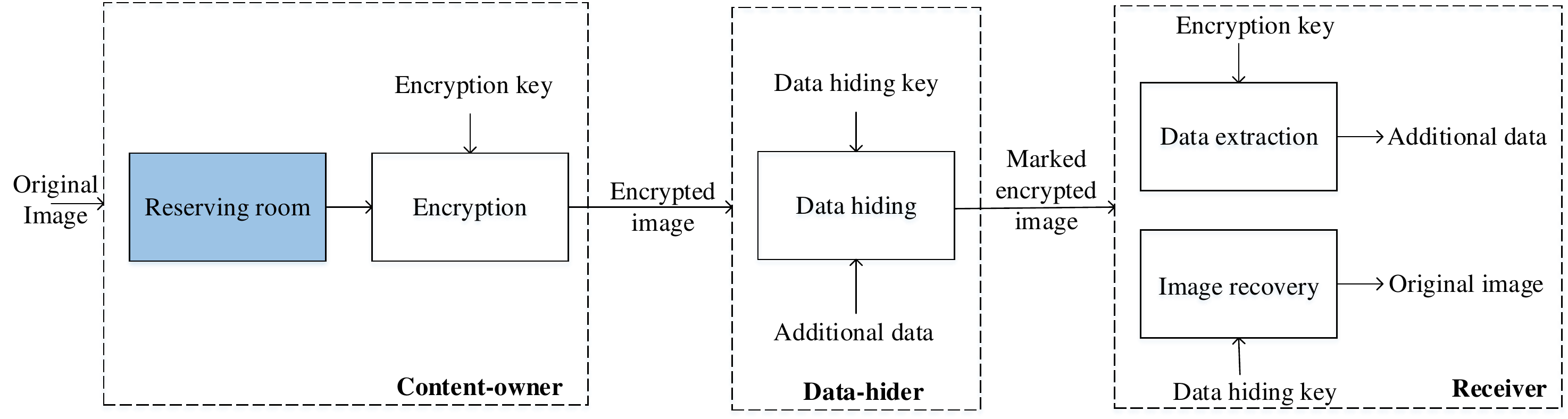}
  }
  \caption{Two categorized frameworks of RDHEI methods: (a) VRAE and (b) RRBE.}
  \label{fig1}
\end{figure*}

 With the growing development of cloud storage and the ever-increasing demand for privacy protection, original images are typically encrypted before they are transmitted. Therefore, reversible data hiding in encrypted images (RDHEI) methods~\cite{puech2008reversible,zhang2014reversibility,xu2016separable} have attracted wide concern over the past years. As shown in Fig.~\ref{fig1}, there are three end users of a RDHEI method: the content-owner, data-hider and receiver. The first end user, content-owner, is the original image provider, who has the right to access the original image and encrypt it. The second end user is the data-hider, who can only receive the encrypted image and embed additional data into the encrypted domain. Besides, the data-hider cannot access the content of original images. The final end user called receiver, he can restore the original image and extract the embedded data. Since the content of original images cannot be obtained after encryption, the visual quality of marked images does not need to be compared. Therefore, embedding capacity is the key metric of RDHEI rather than visual quality.
\par According to the encryption order, the existing RDHEI methods can be divided into two categories, Vacating Room After Encryption (VRAE) and Reserving Room Before Encryption (RRBE), as shown in Fig.~\ref{fig1}. In VRAE methods~\cite{zhang2011reversible,zhang2011separable,hong2012improved,wu2014high,yin2014separable}, data-hiders embed additional data by modifying encrypted pixel values. However, because the entropy of encrypted images has been maximized, VRAE methods can only achieve low embedding capacity. On the premise of ensuring the security of the original image and embedded data, a higher embedding capacity is more applicable to actual problems such as the storage and transmission of medical images, military images and so on. Therefore, to further improve embedding capacity, many RRBE methods have been proposed in the past few years~\cite{ma2013reversible,yi2017binary,puteaux2018efficient,puyang2018reversible,yin2019reversible,wu2019improved,ren2019reversible}. Different from VRAE methods, these methods improve embedding capacity based on the spatial correlation of original images. In this paper, we focus on RRBE methods.
\par In~\cite{ma2013reversible}, Ma et al. first proposed the RRBE method. The key idea of their method is to reserve room by embedding the least significant bit (LSB) of some original pixels into other original pixels. Then generate an encrypted image and embed additional data into the reserved room. In comparison with previous VRAE methods, this method further improves the embedding capacity, and hence makes RRBE method viable. Yi et al.~\cite{yi2017binary} suggested embedding the lower bit planes of original images into the higher bit planes so that the room of lower bit planes can be used to embed data. In the early work, most of methods to increase embedding capacity of RDHEI was to research regarding the LSB substitution, and little attention was paid to the substitution of MSB. Puteaux et al.~\cite{puteaux2018efficient} proposed a new RDHEI method based on MSB substitution instead of LSB, which can achieve higher capacity than previous methods. They designed a MSB prediction method, and generated a label map from which data can be embedded. However, one point of this method can be improved is that only one-MSB can be used. Based on~\cite{puteaux2018efficient}, Puyang et al.~\cite{puyang2018reversible} proposed an extension method using two-MSB prediction, which is more efficient and the performance is improved. Later, Yin et al.~\cite{yin2019reversible} also proposed an improved method based on~\cite{puyang2018reversible}. During their reserving embeddable room phase, multi-MSB of original pixels are predicted adaptively and marked by Huffman coding. According to their results, the embedding capacity is higher than previous methods. In~\cite{wu2019improved}, Wu et al. suggested taking advantage of the spatial correlation of original images and using parametric binary tree labeling to mark image pixels in two different categories. By using this method, original images can be recovered losslessly and embedding capacity can also be improved. Ren et al.~\cite{ren2019reversible} proposed a RDH method for encrypted binary images, which takes advantage of the point that binary images have only two kinds of pixels. Then, additional data is embedded into pixel blocks based on the spatial correlation in this method.
\begin{figure*}[!ht]
  \centering
  \includegraphics[width=1\textwidth]{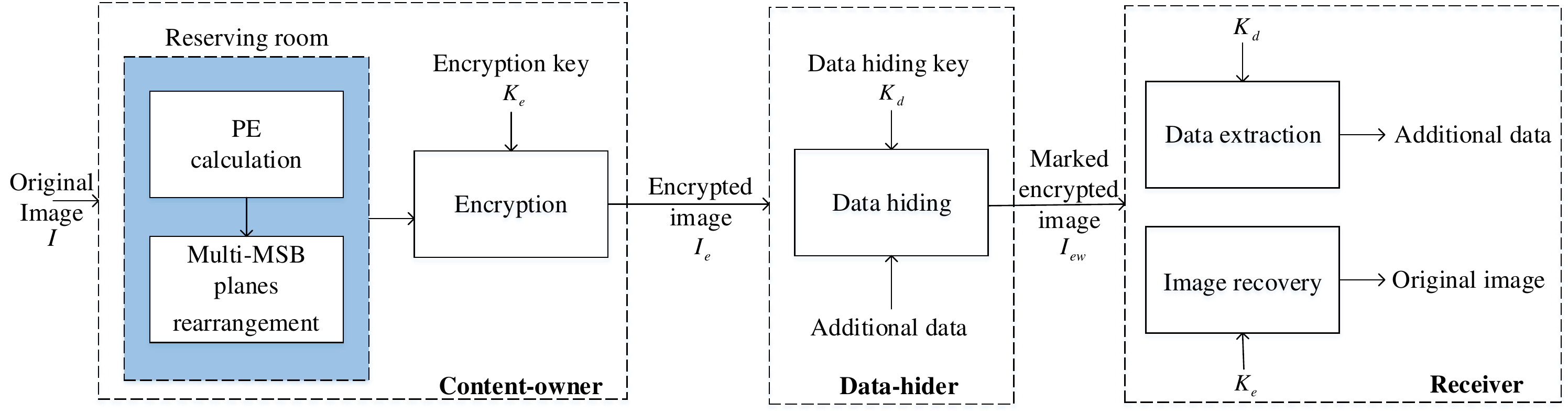}
  \caption{The framework of the proposed method.}
  \label{fig2}
\end{figure*}
\par Although the RDHEI methods mentioned above have achieved relatively good performance, high embedding capacity or full reversibility has not been achieved. Hence, these methods are not satisfactory enough. For example, since Puteaux et al.'s method~\cite{puteaux2018efficient} is limited to MSB and other bits are not fully utilized, the embedding capacity is far from satisfaction. Then, although Yin et al.~\cite{yin2019reversible} have taken into account multiple MSBs, their method generates a lot of auxiliary data. The auxiliary data cannot be compressed well and takes too much room, which makes the net payload of this method not high enough. Wu et al.~\cite{wu2019improved} proposed to mark pixels with a parameter binary tree and divide them into two categories, which would make some pixels unable to embed data and affect the performance of embedding capacity. Considering the similarity between bit planes of grayscale images and binary images, that is, both bit planes and binary images are only composed of two types of values. Ren et al.'s scheme~\cite{ren2019reversible} was directly applied to grayscale images by us, however, the embedding capacity is not satisfactory when this application is reversible.
There are two main reasons for the limited embedding capacity: In the application, the bit planes of original images don't have enough room to accommodate additional data compared with the prediction error (PE) bit planes in our paper; This application will generate too much auxiliary data without significant sparse features, hence it cannot be well compressed. To achieve better performance, in this paper, we propose a new method which effectively reduces these unsatisfactory points.
\par The main contributions of our method can be summarized as follows:
\par 1) We successfully extend the pixel prediction scheme of binary image to bit planes of grayscale image, and solve the problem that high embedding capacity and reversibility cannot coexist. In a word, after we improve the embedding capacity, our method is still reversible.
\par 2) We make full use of the correlation of adjacent pixels and compress auxiliary data with sparse features via arithmetic coding, which makes more embeddable positions in the bit planes. Experimental results also prove that we have achieved a higher embedding capacity compared with the most advanced RDHEI method.
\par 3) Furthermore, we have successfully used the median edge predictor (MED). The distribution of PE values becomes smaller relative to the distribution range of original pixel values, which exploits spatial redundancy better and makes the available embedded room larger.
\par The rest of this paper is organized as follows: The proposed method is described in detail in Section~\ref{sec::Proposed method}, and Section~\ref{sec::Experimental} introduces the experimental results and analysis of our method. Finally, this paper is concluded in Section~\ref{sec::Conclusion}.

\section{Proposed method}
\label{sec::Proposed method}
In this section, a high-capacity and fully reversible RDHEI method based on pixel prediction and multi-MSB planes rearrangement is proposed. As shown in Fig.~\ref{fig2}, the proposed method can be divided into three parts: 1) Embeddable room reservation and encryption are done by content-owner, 2) Data embedding is done by data-hider, 3) Data extraction and image recovery are done by receiver. In this section, the proposed method is introduced in detail. First, Section~\ref{subsec::room reservation} describes the process of reserving embeddable room. Next, in Section~\ref{subsec::encrypted image}, procedures of image encryption are given. Then, Section~\ref{subsec::marked encrypted image} presents the process of additional data embedding. In the last Section~\ref{subsec::extraction}, data extraction and image recovery are described in detail.
\subsection{Embeddable room reservation}
\label{subsec::room reservation}
\par The procedures of this section are divided into three parts: Calculation of predicted errors, generation of bit planes and rearrangement of bit planes.
\subsubsection{Calculation of prediction errors}
\par
\begin{figure}[htbp]
  \centering
  \includegraphics[width=0.1\textwidth]{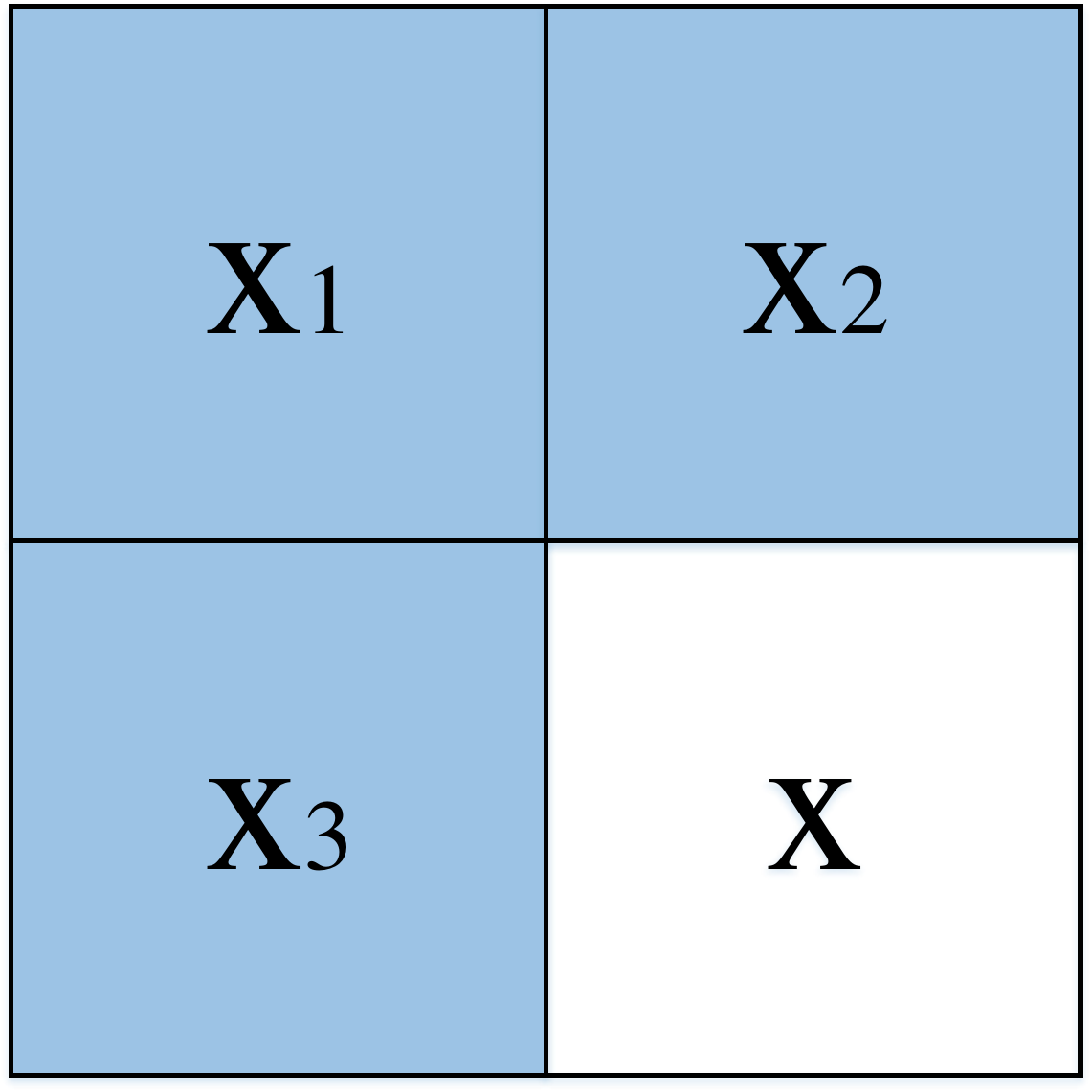}
  \caption{The context of the current pixel by MED predictor.}
  \label{fig3}
\end{figure}

 For an original image ${ I }$ sized ${ M \times N }$, predicted value ${ px }$ is calculated by the MED predictor~\cite{weinberger2000loco}. As shown in Fig.~\ref{fig3}, ${ px }$ is calculated based on the three pixels around the current pixel ${ x }$. In particular, the pixels of first row and first column will be the reference pixels to recover original image, hence the value of them will not be changed.
\par The detailed  calculation formula of predicted values is as follows:
\begin{equation}
  \label{eq::eq1}
  \centering
  px = \left\{\begin{matrix}
    {\rm max}(x_{2},x_{3})  & , & x_{1}\leq {\rm min}(x_{2},x_{3}) \\
    {\rm min}(x_{2},x_{3})  & , & x_{1}\geq {\rm max}(x_{2},x_{3}) \\
    x_{2}+x_{3}-x_{1} & , & \rm otherwise
  \end{matrix}\right..
\end{equation}

\par Then, the PE ${ e(i,j) }$ can be obtained as follows:
\begin{equation}
  \label{eq::eq2}
  \centering
  e(i,j) = \left\{\begin{matrix}
    x(i,j)         & , & i=1\;{\rm or}\;j=1 \\
    x(i.j)-px(i,j) & , & \rm otherwise
  \end{matrix}\right.,
\end{equation}
\noindent where ${1\leq i \leq M , 1\leq j \leq N }$.
\par Finally, we define ${ e(i,j) }$ greater than 64 or less than -64 as overflow pixels, and use Eq.~\eqref{eq::eq3} to change the value of these pixels. The reason for this operation will be explained later. After the calculation, the distribution of PEs is more concentrated than that of the original pixel values, which means that there are more identical values in bit planes and the embedding capacity will be improved.
\begin{equation}
  \label{eq::eq3}
  \centering
  e(i,j)=x(i,j)\quad,\quad e(i,j)<-64\;{\rm or}\;e(i,j)>64.
\end{equation}
\subsubsection{Generation of bit planes}
\par After calculating the PEs, the PE bit planes are generated. If only traditional method, Eq.~\eqref{eq::eq4}, is used to calculate bit planes, the proposed method will not be reversible. Since the original positions of bit planes are shuffled in the subsequent embedding operation, whether the PE is positive or negative during the image recovery step cannot be determined. Based on this problem, an adaptive method is proposed to calculate PE bit-planes by us. The method is introduced in detail as follows: First, considering the range of pixel values of grayscale images, the PEs of reference pixels are converted into 8-bit binary sequences by Eq.~\eqref{eq::eq4}, where ${ i }$=1 or ${ j }$=1 and ${ \left \lfloor * \right \rfloor }$ is the floor operations.
\begin{equation}
  \label{eq::eq4}
  \centering
  e^{k}(i,j)=\left \lfloor \frac{e(i,j) \bmod 2^{9-k}}{2^{8-k}} \right \rfloor,\, k=1,2,\dots,8.
\end{equation}
\par Next, other PEs are converted into 7-bit binary sequences by Eq.~\eqref{eq::eq5}. Since there are some overflow PEs, we use Eq.~\eqref{eq::eq6} to calculate the eighth binary bit of PEs in different situations. Specifically, signs of non-overflow PEs are represented by one bit plane and absolute values of non-overflow PEs are represented by other bit planes. Moreover, only a few PE values are overflow values (the PE values less than - 64 or greater than 64 are overflow PEs), hence the final reconstructed image is still lossless. The calculation formulas for this step are as follows:
\begin{equation}
  \label{eq::eq5}
  \centering
  e^{k}(i,j)=\left \lfloor \frac{e(i,j) \bmod 2^{9-k}}{2^{8-k}} \right \rfloor,\, k=1,2,\dots,7,
\end{equation}
\begin{equation}
  \label{eq::eq6}
  \centering
  e^{8}(i,j) = \left\{\begin{matrix}
    1- \textrm{sign}\,e(i,j) & , & -64\leq e(i,j)\leq64 \\
    e(i,j) \bmod 2           & , & \rm otherwise            \\
  \end{matrix}\right..
\end{equation}
\begin{figure}[!ht]
  \centering
  \includegraphics[width=0.45\textwidth]{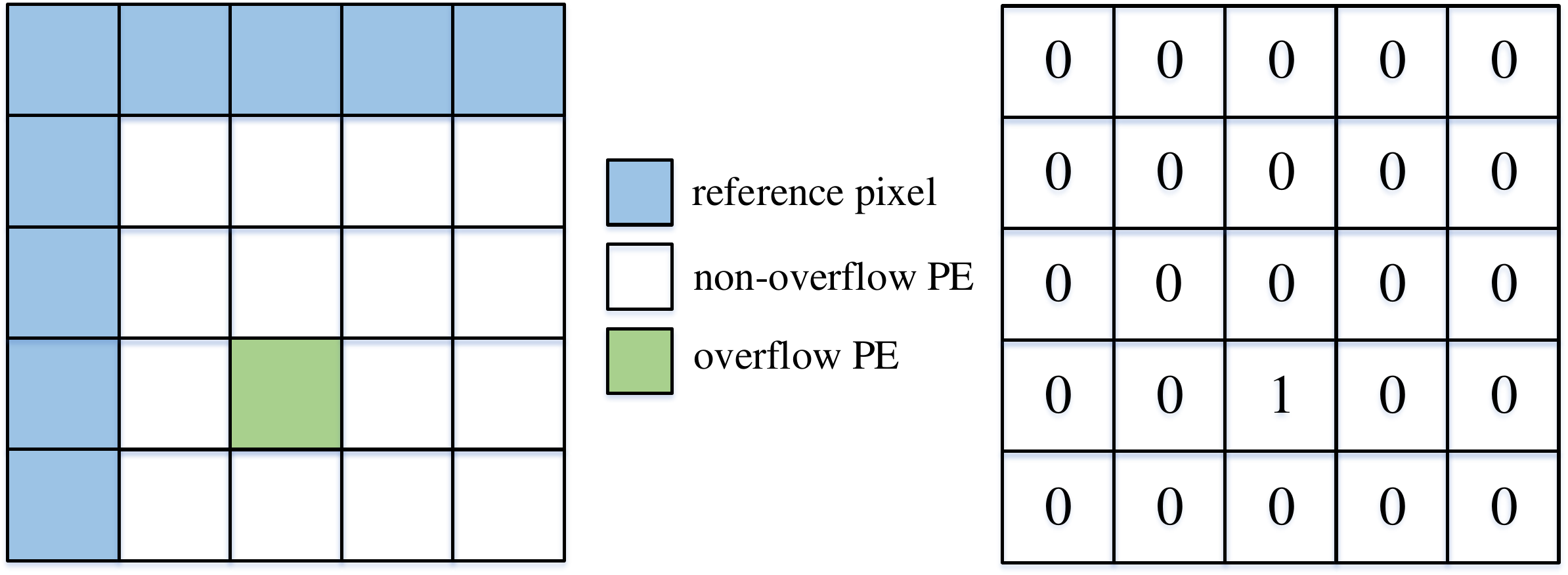}
  \caption{The label map of PE image.}
  \label{fig4}
\end{figure}
\par Finally, because PEs are calculated by different formulas, a PE label map must be generated to identify each PE. We set the tag 1 for overflow PEs and 0 for other PEs. In this way, a label map ${ L_1 }$ with a large number of 0 and a fairly small number of 1 is generated. Particularly, ${ L_1 }$ can be well compressed by arithmetic coding and only occupies a very small room. For better understanding, as shown in Fig.~\ref{fig4}, an example is given to describe our method.
\subsubsection{Rearrangement of bit planes}
\par
\begin{figure}[!ht]
  \centering
  \includegraphics[width=0.4\textwidth]{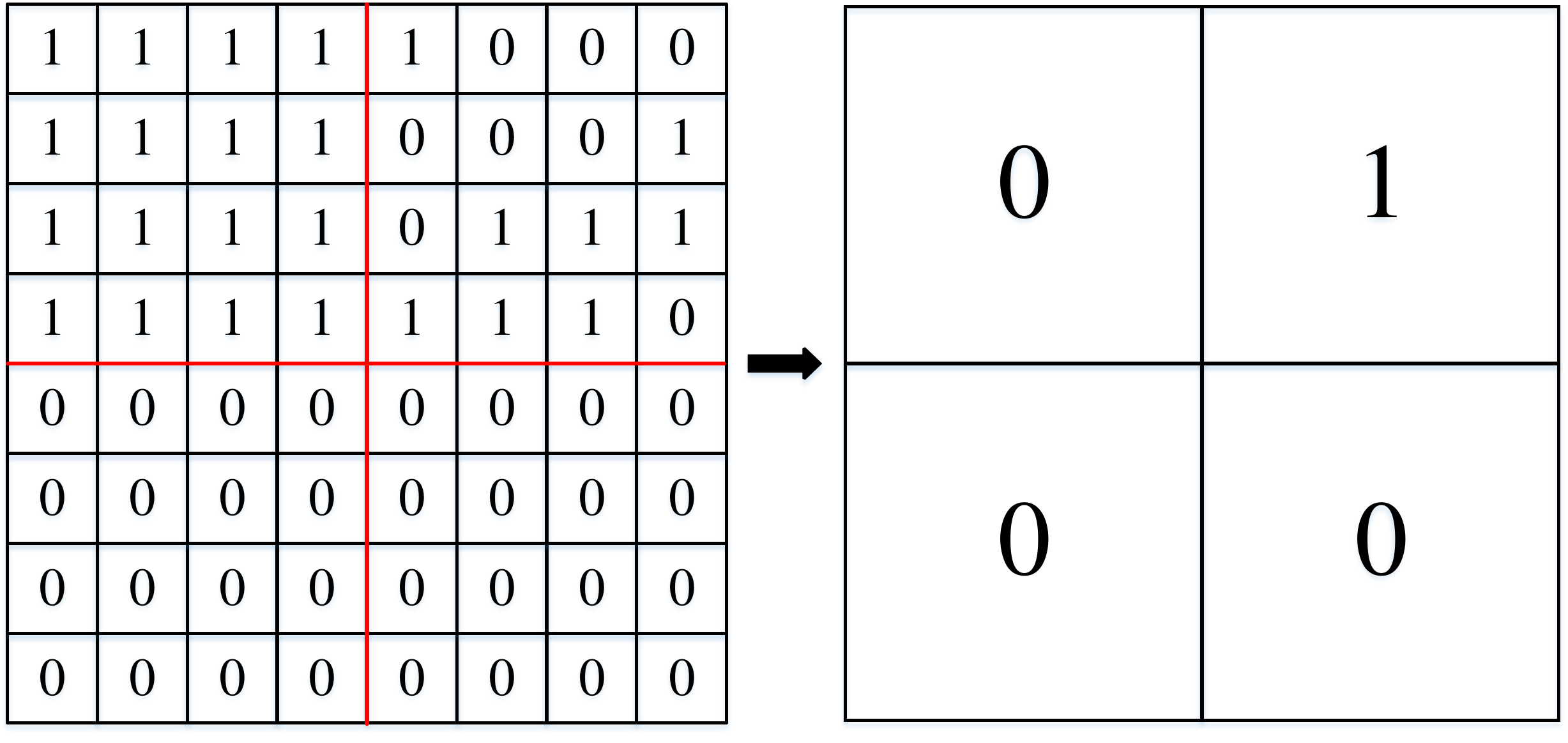}
  \caption{An example of original blocks labeling of bit planes.}
  \label{fig5}
\end{figure}

\begin{figure}[!ht]
  \centering
  \includegraphics[width=0.4\textwidth]{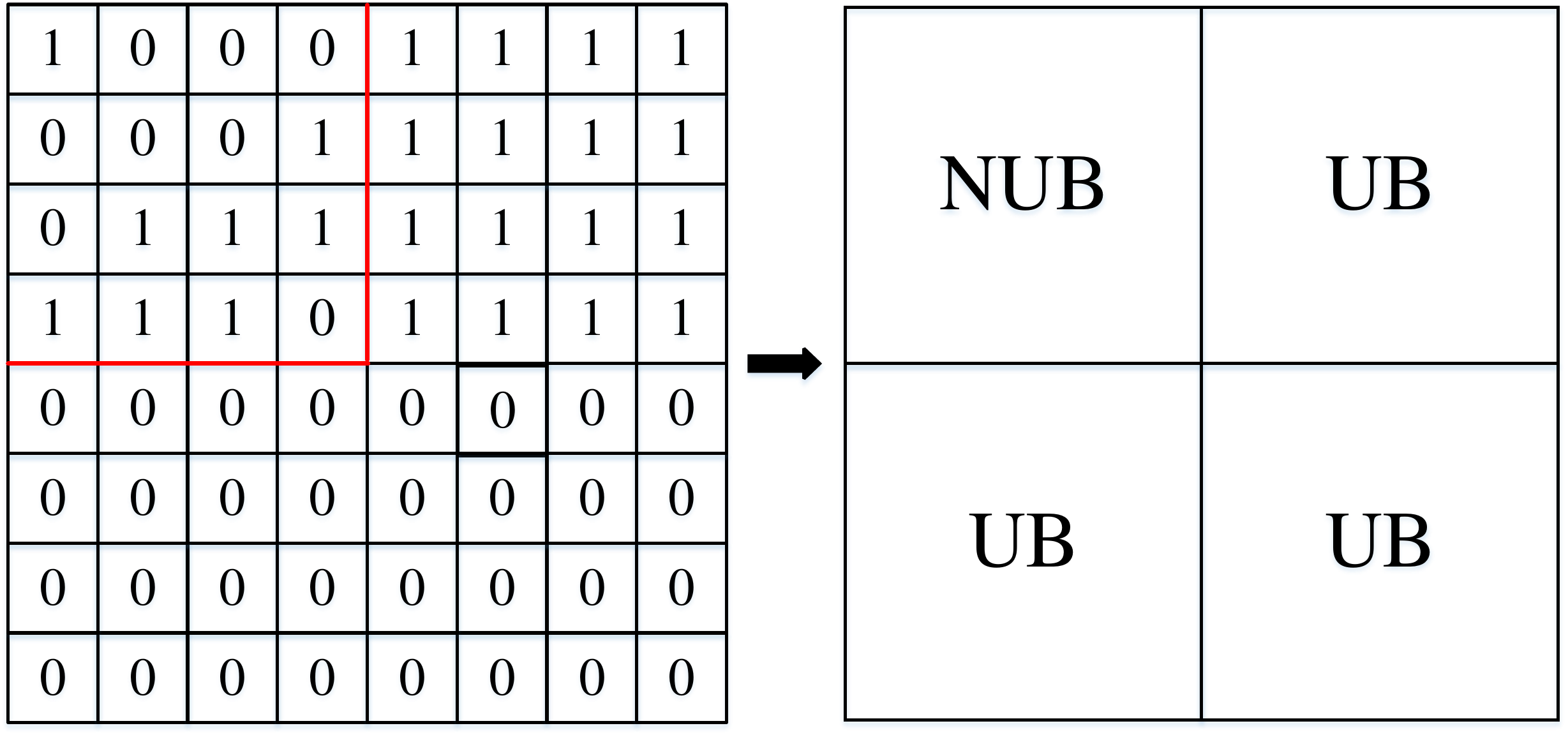}
  \caption{Rearrangement of bit planes.}
  \label{fig6}
\end{figure}

 As shown in Fig.~\ref{fig5}, the bit planes obtained by the above steps are divided into several non-overlapping blocks sized ${ k \times k }$. All values are the same for uniform blocks (UB), otherwise they are non-uniform blocks (NUB). Since there are much redundancy in the multi-MSB planes, the number of UB is much larger than that of NUB. Based on this feature, NUB and UB are labeled by 1 and 0 respectively. The label map ${ L_2 }$ with sparse feature can be obtained and well compressed, and the length of the map is expressed in a 16-bit binary sequence. Then, the processed bit planes are traversed in order, the NUBs of each bit plane are arranged in order in the upper order, and the UBs in the lower order, as shown in Fig.~\ref{fig6}. Meanwhile,all of NUBs need to be marked, and the NUB that can be embedded into additional data is marked with 0, otherwise 1. The specific basis for judging whether NUB can be embedded will be mentioned later.
\par Lastly, we embed all the auxiliary data into UBs of the corresponding bit plane, and the last value of each UB is used as the predicted value without auxiliary data. Moreover, since some bit planes are not enough to accommodate auxiliary data, an 8-bit binary sequence is finally generated to indicate whether the current bit plane can be embedded.

\begin{figure*}[!ht]
  \centering
  \includegraphics[width=0.9\textwidth]{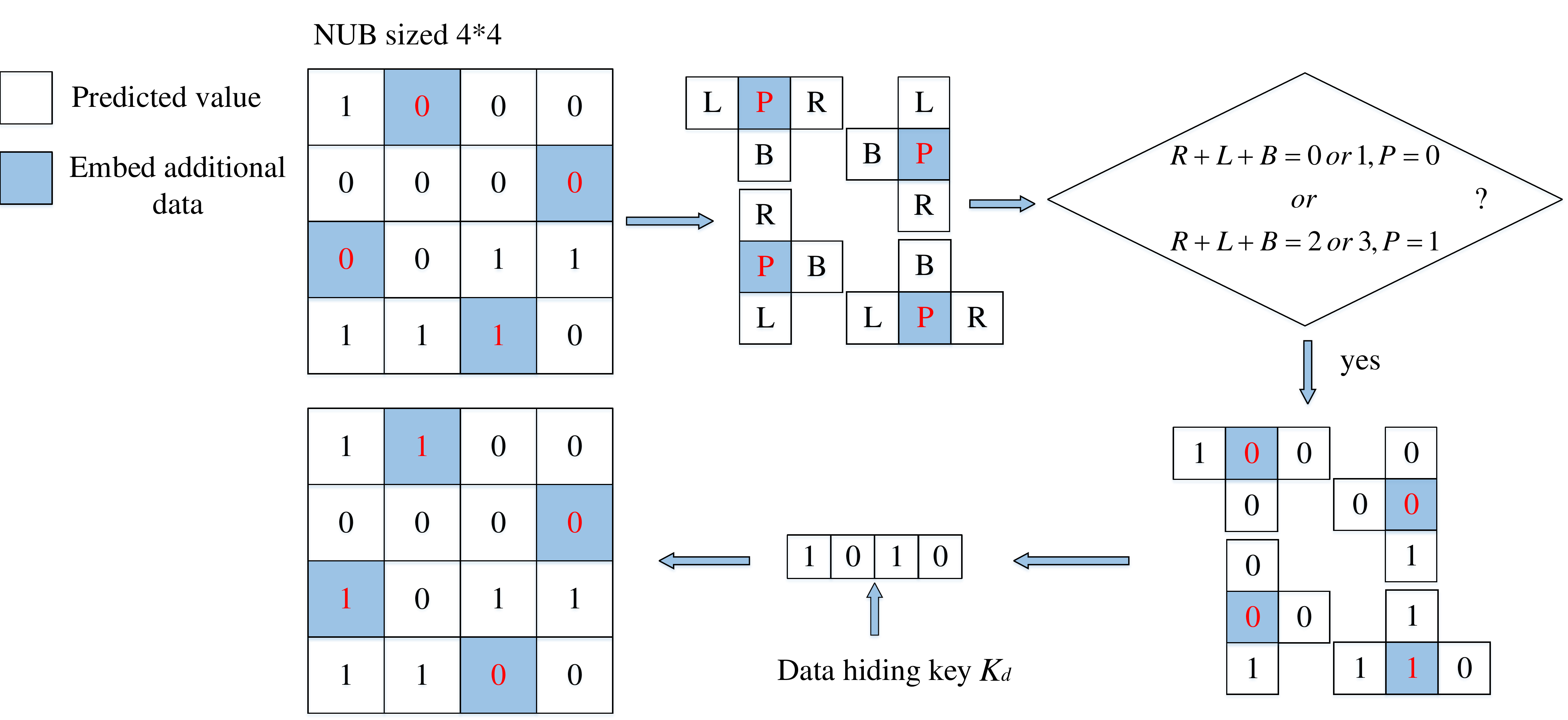}
  \caption{An example of embedding data into NUB.}
  \label{fig7}
\end{figure*}
\subsection{Generation of the encrypted image}
\label{subsec::encrypted image}
\par After reserving the embeddable room, the image is divided into eight bit planes, and the auxiliary data is extracted sequentially in the lower right of each bit plane in order to locate the encrypted position. Then, a pseudo-random matrix ${ H }$ of size ${ M \times N }$ is generated by an encryption key ${ K_{e} }$, the values of this matrix are converted into 8-bit binary sequences by Eq.~\eqref{eq::eq7},
\begin{equation}
  \label{eq::eq7}
  \centering
  H^{k}(i,j)=\left \lfloor \frac{H(i,j) \bmod 2^{9-k}}{2^{8-k}} \right \rfloor,\, k=1,2,\dots,8,
\end{equation}
\noindent where ${1\leq i\leq M, 1\leq j\leq N}$. At last, for the original PE ${ e^{k}(i,j) }$ of each bit plane that can be encrypted, we use the Eq.~\eqref{eq::eq8} to perform encryption operations and ${ \oplus }$ denotes exclusive-or (XOR) operation:
\begin{equation}
  \label{eq::eq8}
  \centering
  e_{e}^{k}(i,j)=e^{k}(i,j)\oplus H^{k}(i,j),\; k=1,2,\dots,8.
\end{equation}
\noindent In this way, we can calculate the encrypted PE ${ e_{e}^{k}(i,j) }$ of ${ k^{th} }$ bit plane and finally get the encrypted image ${ I_{e} }$.
\subsection{Generation of the marked encrypted image}
\label{subsec::marked encrypted image}
\par Considering the similarity between bit planes and binary images, the pixel prediction scheme is used to embed additional data into the processed PE bit planes, which is proposed by Ren et al.~\cite{ren2019reversible}. As mentioned before, each processed bit plane is divided into UBs and NUBs. For different blocks, different schemes are used to embed additional data. Besides, to improve the security of our method, the data hiding key ${ K_{d} }$ is used to encrypt additional data before embedding. After completing these steps, embedding steps are described in detail below.
\par Auxiliary data in the lower right corner of the MSB plane needs to be extracted first to determine whether the data can be embedded into the current bit plane and the location of the embedded data. The values of bit planes are only 0 and 1, which means that the probability of three adjacent values around each value being identical is very high. According to pixel correlation, for each NUB, if the value of middle position is 0, the sum of three neighboring values may be 0 or 1 (that is, the probability of neighboring values being 0 is higher than 1). Similarly, if the middle value is 1, the sum of three neighboring values may be 2 or 3. Based on this feature, we judge each NUB, if current NUB satisfies the feature, it is an embeddable block. Conversely, additional data cannot be embedded. Fig.~\ref{fig7} shows an example of embedding data into NUB. Each NUB is divided into four parts, and each part is composed of four values, where ${ P }$ is the middle position of each part that can be embedded, and ${ L }$, ${ B }$, ${ R }$ are the three adjacent predicted values around ${ P }$.
\begin{figure}[!ht]
  \centering
  \includegraphics[width=0.45\textwidth]{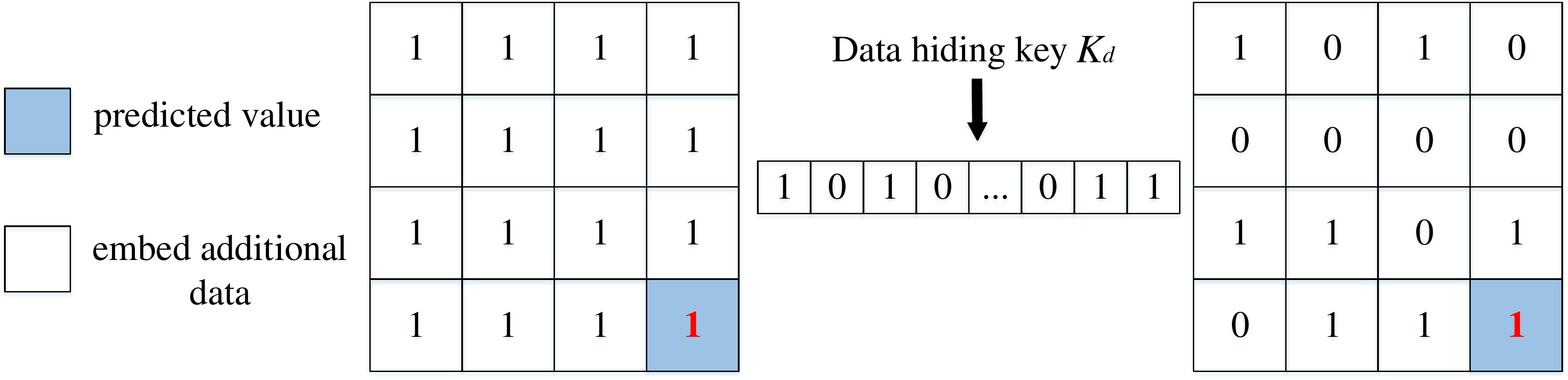}
  \caption{An example of embedding data into UB.}
  \label{fig8}
\end{figure}
\par As shown in Fig.~\ref{fig8}, for each UB, the values of one block are identical. It means that, for a UB, we only need to keep one of the value in the block not being changed, then all values can be recovered in the block after taking the unchanged value as the prediction value. Based on this characteristic, the additional data is embedded into the UB except the position of predicted value. To facilitate understanding, here is an example of embedding data into a UB. As shown in Fig.~\ref{fig8}, after the data is embedded into the UB, only the last value remains unchanged. Finally, after embedding additional data into NUBs and UBs, we have generated the marked encrypted image ${ I_{ew} }$.
\begin{figure*}[!ht]
  \centering
  \begin{minipage}[t]{0.18\textwidth}
    \centering
    \subfigure[]{
      \label{fig-9-a}
      \includegraphics[width=\textwidth]{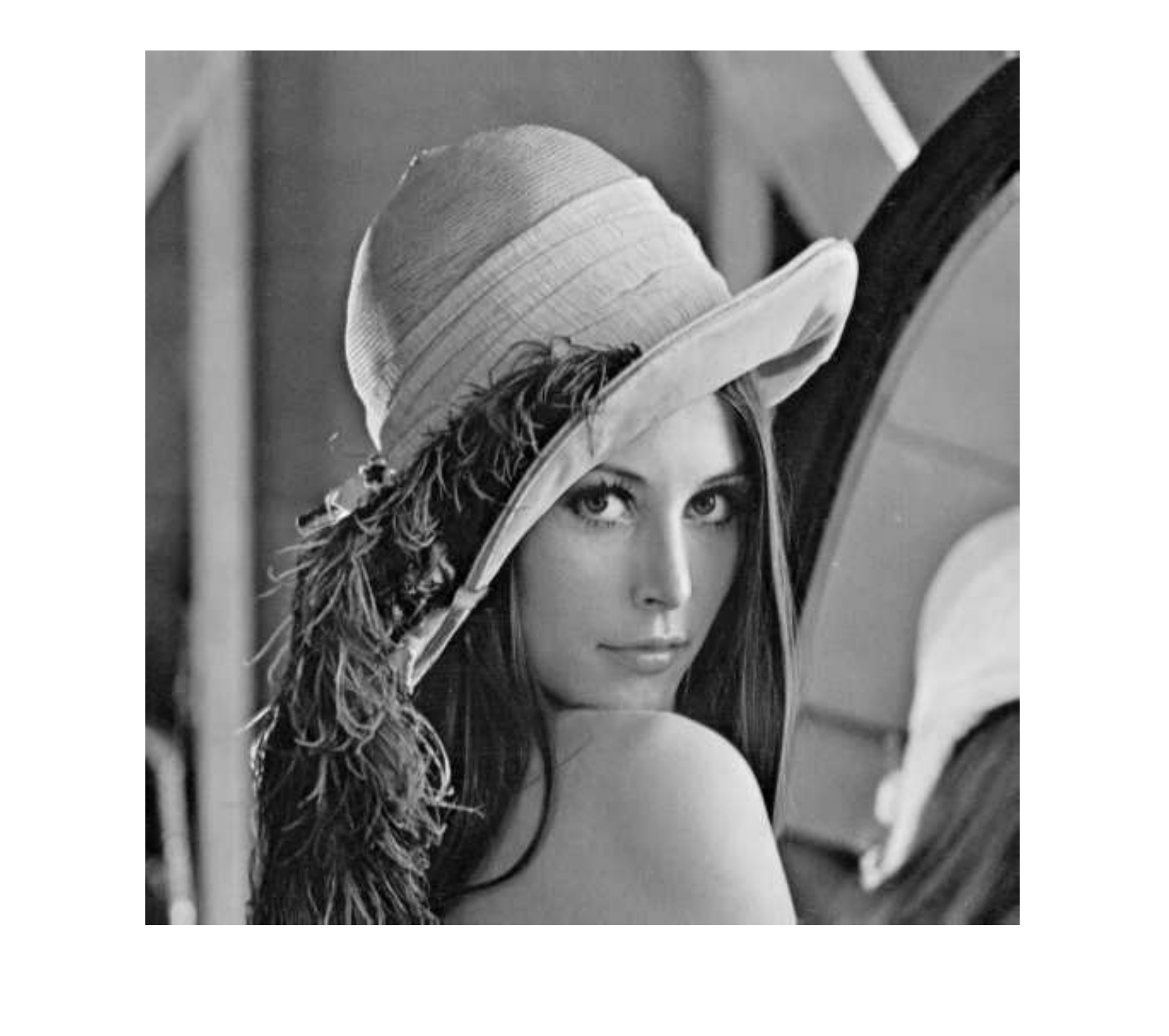}
    }
  \end{minipage}
  \begin{minipage}[t]{0.18\textwidth}
    \centering
    \subfigure[]{
      \label{fig-9-b}
      \includegraphics[width=\textwidth]{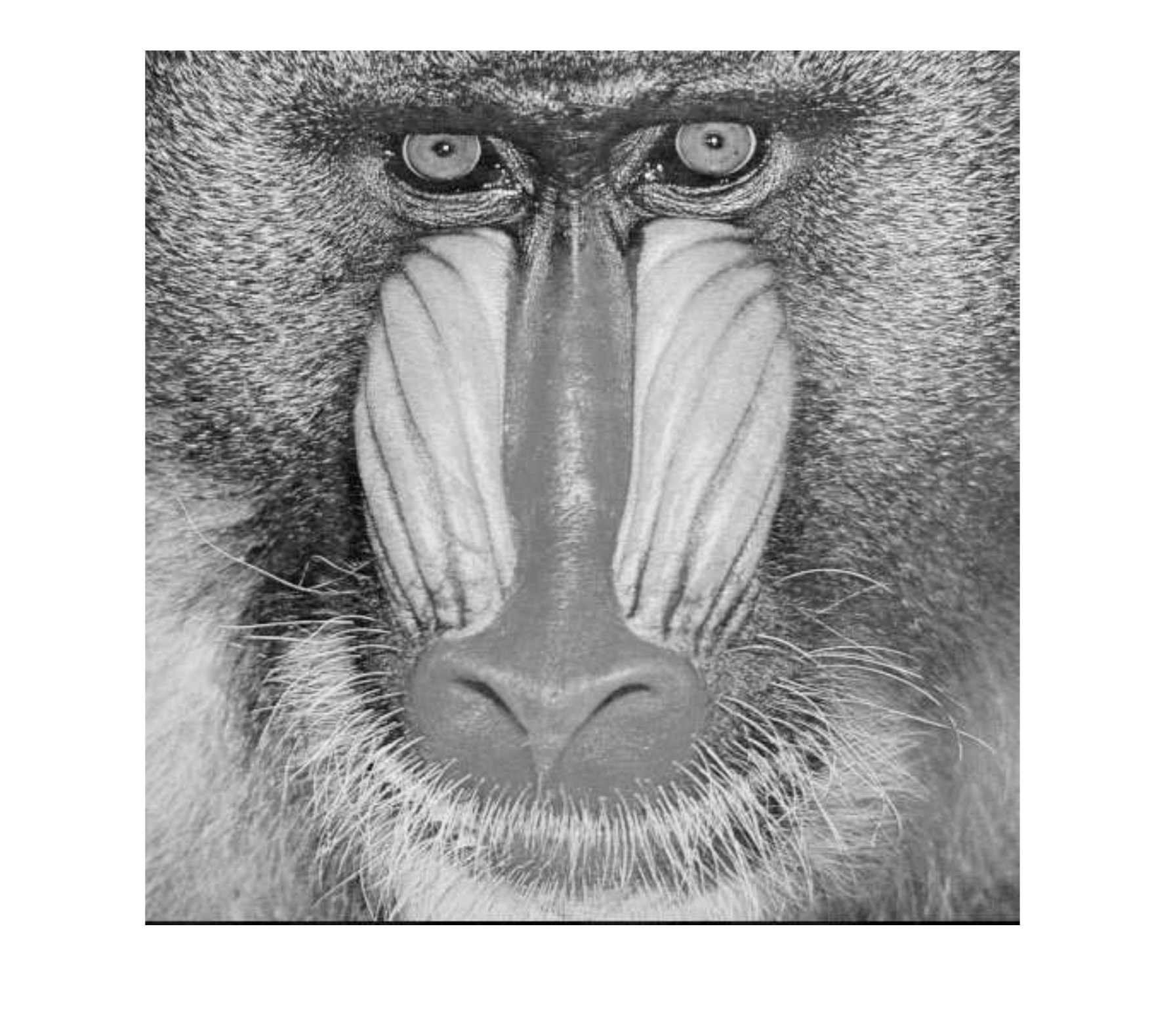}
    }
  \end{minipage}
  \begin{minipage}[t]{0.18\textwidth}
    \centering
    \subfigure[]{
      \label{fig-9-c}
      \includegraphics[width=\textwidth]{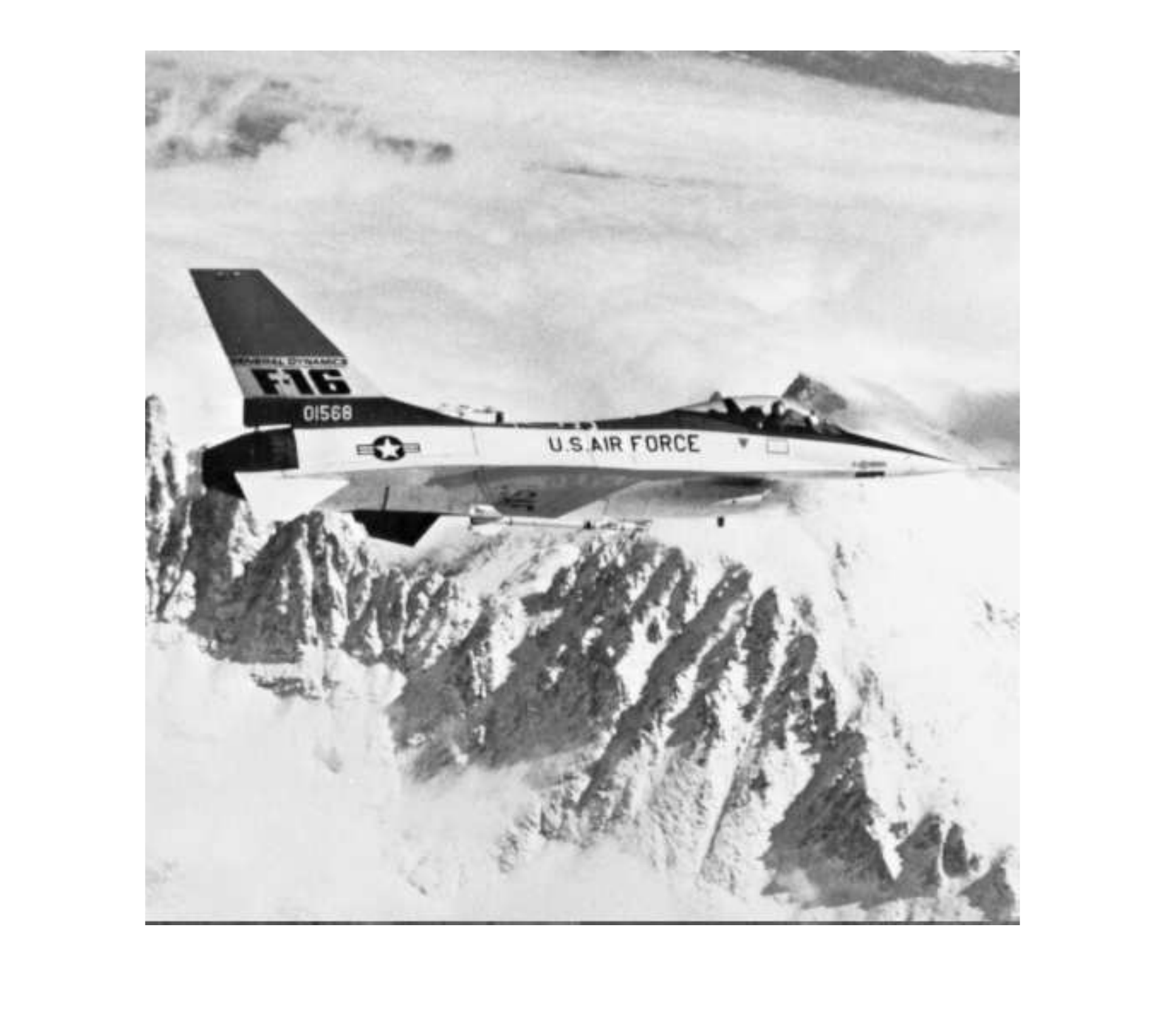}
    }
  \end{minipage}
  \begin{minipage}[t]{0.18\textwidth}
    \centering
    \subfigure[]{
      \label{fig-9-d}
      \includegraphics[width=\textwidth]{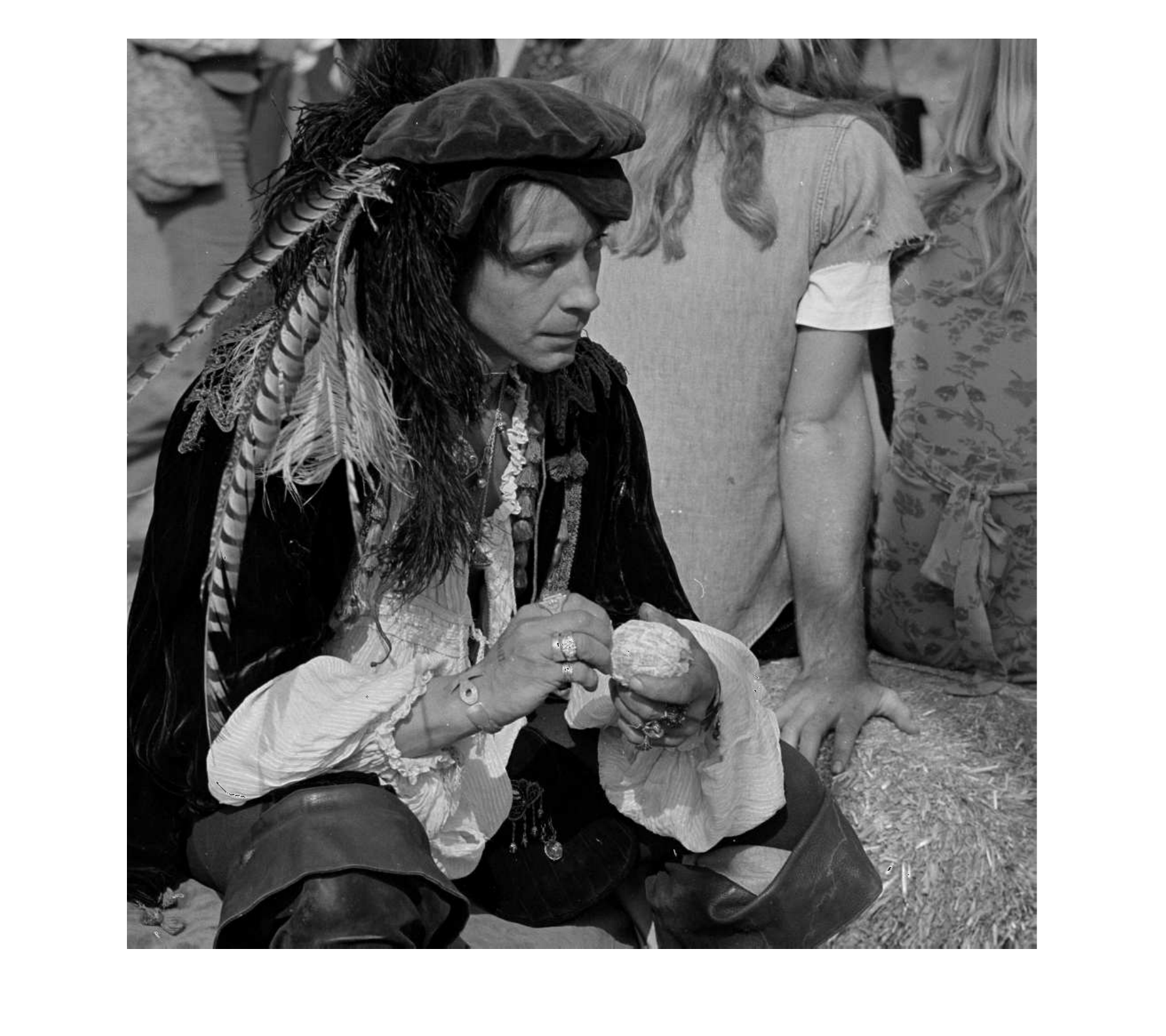}
    }
  \end{minipage}
  \begin{minipage}[t]{0.18\textwidth}
    \centering
    \subfigure[]{
      \label{fig-9-e}
      \includegraphics[width=\textwidth]{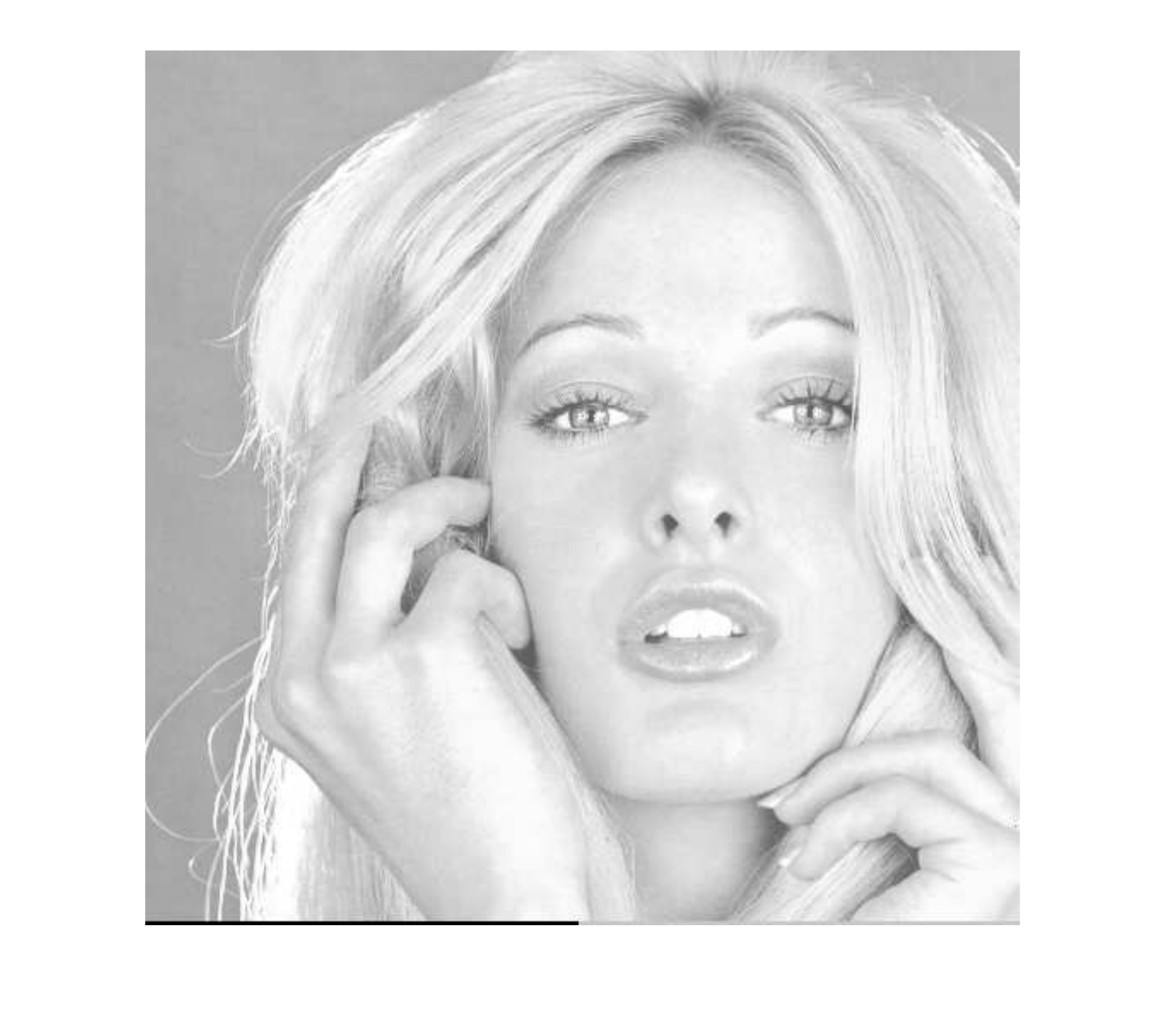}
    }
  \end{minipage}
  \caption{ Test images: (a) \emph{Lena}; (b) \emph{Baboon}; (c) \emph{Jetplane}; (d) \emph{Man}; (e) \emph{Tiffany}.}
  \label{fig9}
\end{figure*}

\begin{figure*}[!ht]
  \centering
  \begin{minipage}[t]{0.18\textwidth}
    \centering
    \subfigure[]{
      \label{fig-10-a}
      \includegraphics[width=\textwidth]{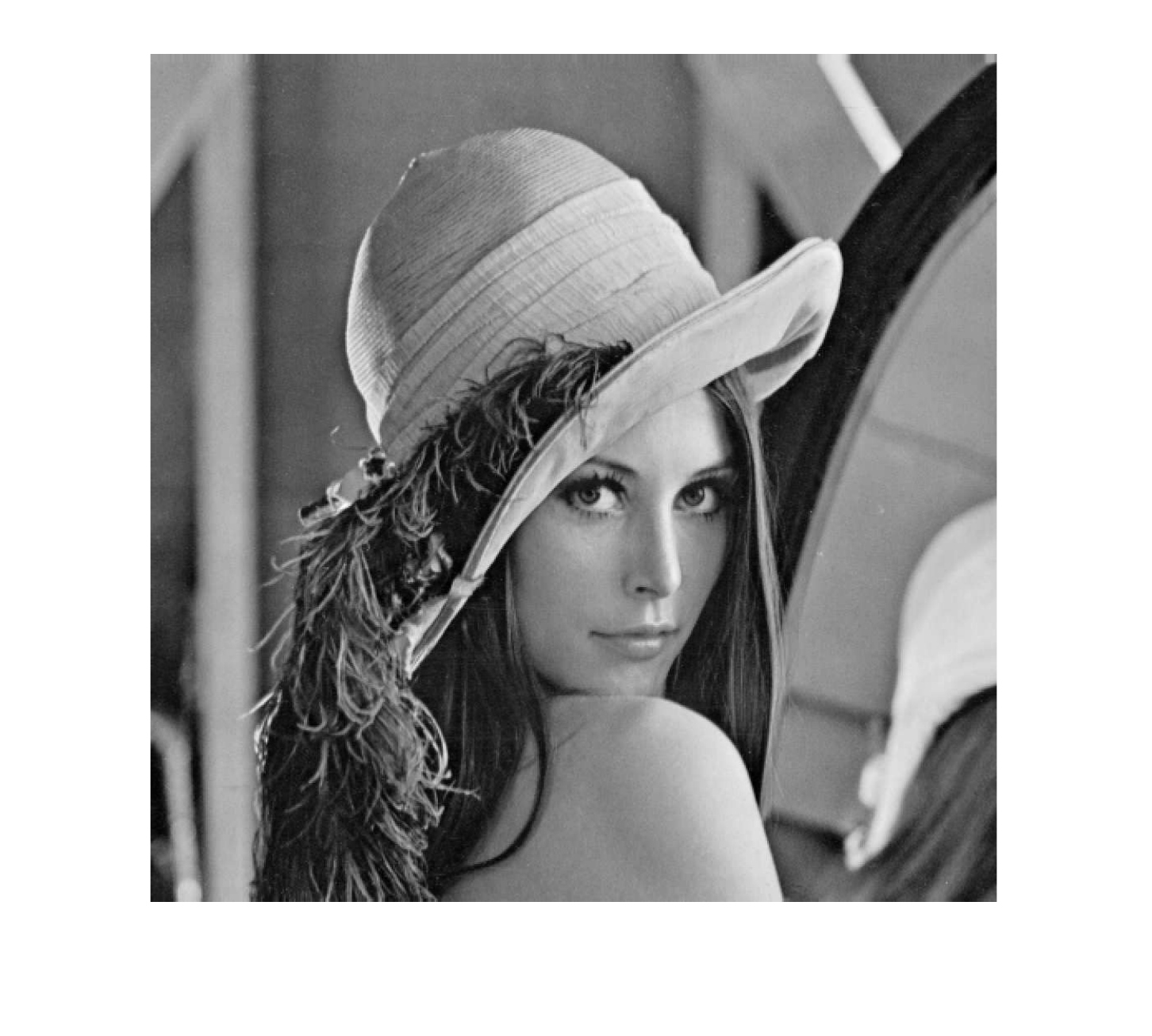}
    }
  \end{minipage}
  \begin{minipage}[t]{0.18\textwidth}
    \centering
    \subfigure[]{
      \label{fig-10-b}
      \includegraphics[width=\textwidth]{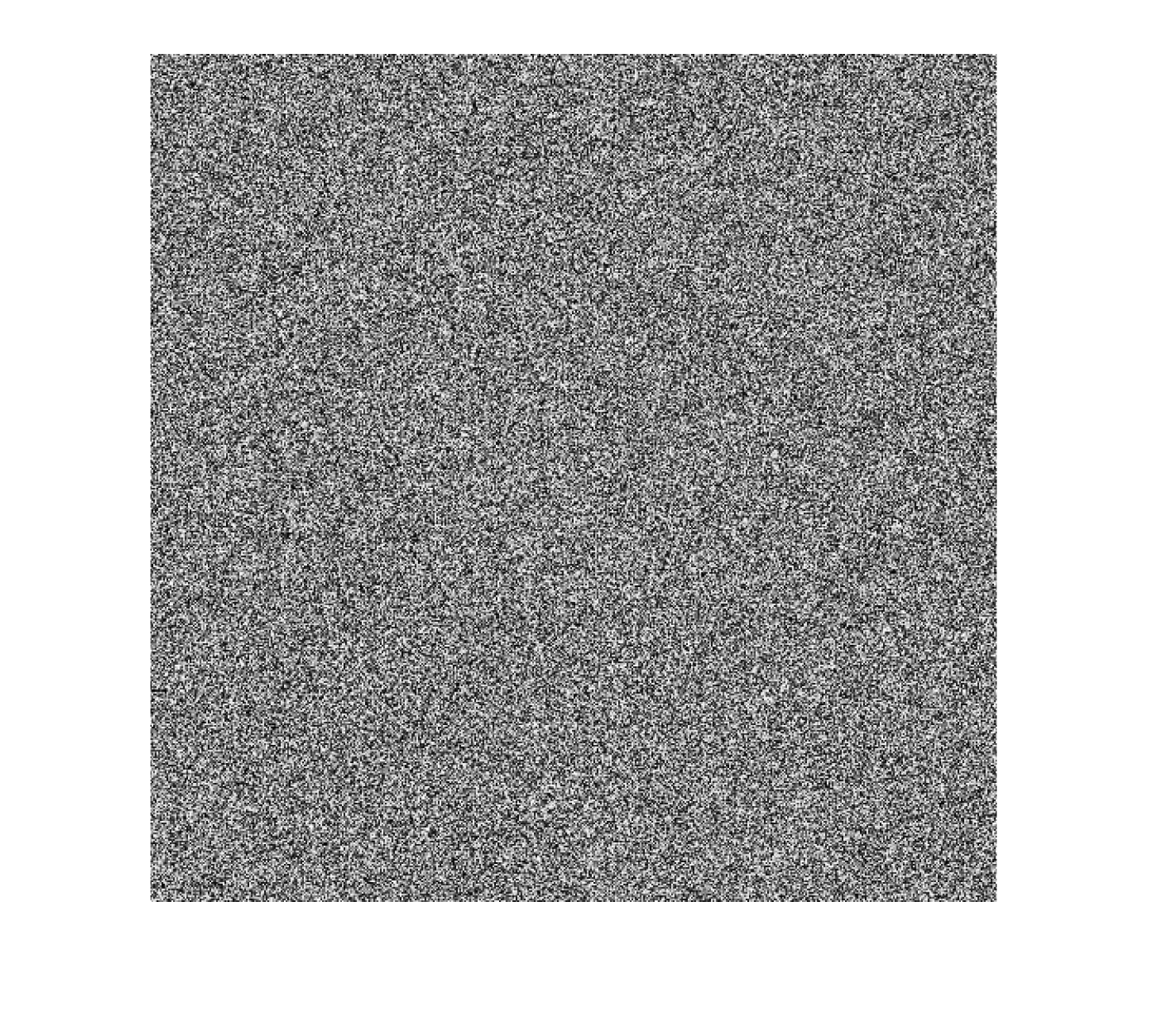}
    }
  \end{minipage}
  \begin{minipage}[t]{0.18\textwidth}
    \centering
    \subfigure[]{
      \label{fig-10-c}
      \includegraphics[width=\textwidth]{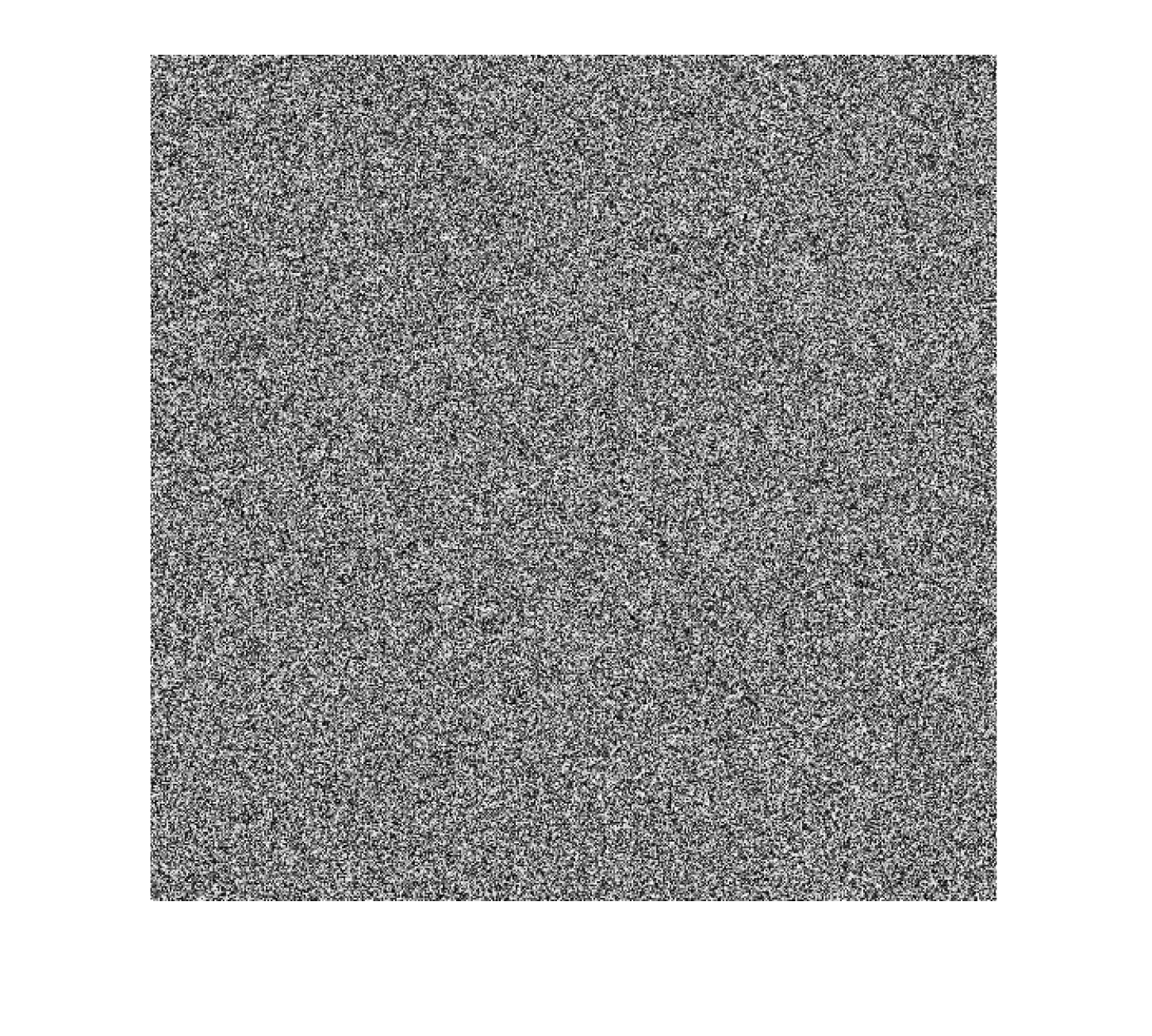}
    }
  \end{minipage}
  \begin{minipage}[t]{0.18\textwidth}
    \centering
    \subfigure[]{
      \label{fig-10-d}
      \includegraphics[width=\textwidth]{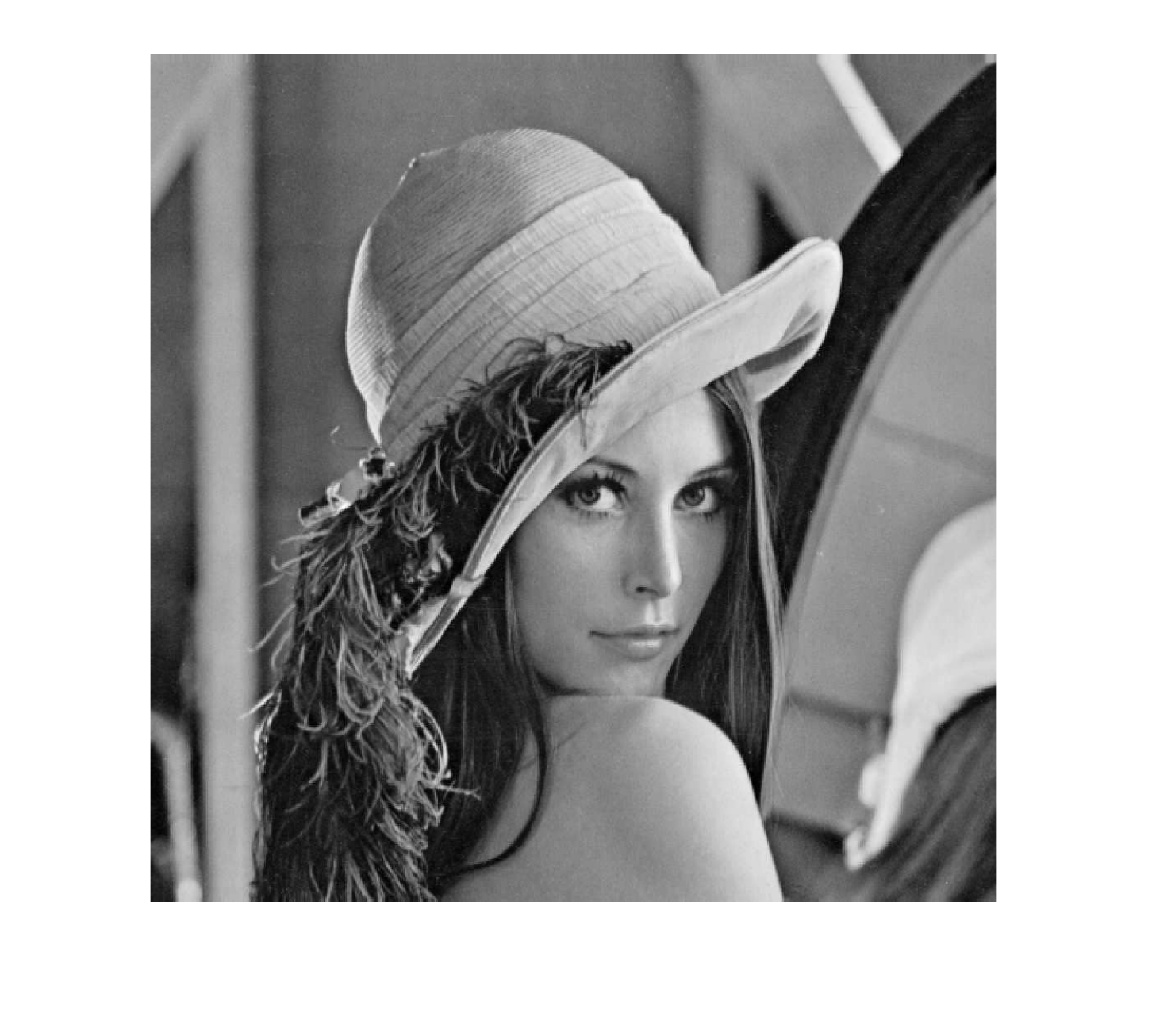}
    }
  \end{minipage}
  \caption{ Results of applying our method to \emph{Lena} image when ${ k }$=4: (a) Original image ${ I }$; (b) Encrypted image ${ I_{e} }$; (c) Marked encrypted image ${ I_{ew} }$, with net payload ER = 2.87 ${ bpp }$; (d) Reconstructed image ${ I }$. }
  \label{fig10}
\end{figure*}

\subsection{Data extraction and image recovery}
\label{subsec::extraction}
\par Whether the receiver obtains the embedded data or the original image depends on which key the receiver has. Therefore, three cases are given according to the different keys owned by the receiver:
\par 1) If the receiver has the data hiding key ${ K_{d} }$, the embedded data can be obtained without errors. First, the legal receiver divides the marked encrypted image into eight bit planes according to the auxiliary data in the lower right corner of the MSB plane. If the current bit plane is an embeddable bit plane, locate the position where the additional data is embedded, and extract auxiliary data of remaining embeddable bit planes in sequence. For each NUB, if the auxiliary data of this NUB is 0, additional data can be extracted in this NUB. For each UB, except for the prediction pixel of the last position, the additional data of other positions is extracted in order. Finally, all embedded additional data can be extracted without errors and decrypted with the data hiding key ${ K_{d} }$. Since the receiver does not have the encryption key ${ K_{e} }$, the original image cannot be recovered.
\par 2) If the receiver has the encryption key ${ K_{e} }$, the original image can be reconstructed losslessly. First, the legal receiver decrypts the marked encrypted image with the encryption key ${ K_{e} }$, and extracts auxiliary data in the lower right corner of the MSB plane to determine whether the current bit plane is rearranged. Then, the auxiliary data of remaining bit planes is sequentially extracted. According to the prediction scheme mentioned above, all of NUBs and UBs in embeddable bit planes can be restored. Finally, according to the extracted label map ${ L_2 }$, the receiver can reconstruct the original image losslessly.
\par 3) If the receiver has both the data hiding key ${ K_{d} }$ and the encryption key ${ K_{e} }$, then the data can be extracted error-free and images can be recovered losslessly. The detailed steps are the same as above.

\section{Experimental results and analysis}
\label{sec::Experimental}
\par To clearly evaluate the performance of the proposed method, experimental results and analysis are detailed in this section. First, several experimental results are summarized in Section~\ref{subsec::experimental results}. As shown in Fig.~\ref{fig9}, five commonly used test images \emph{Lena}, \emph{Baboon}, \emph{Jetplane}, \emph{Man} and \emph{Tiffany} are used as examples to show the experimental results. Furthermore, we also realize tests on three public datasets: BOSSbase~\cite{bas2011break}, BOWS-2~\cite{bas2017image} and UCID~\cite{schaefer2003ucid}. Then, Section~\ref{subsec::performance analysis} gives performance analysis in terms of reversibility and embedding capacity. Finally, Section~\ref{subsec::Comparison} compares the proposed method with recent state-of-the-art methods.

\subsection{Experimental results}
\label{subsec::experimental results}
\par Fig.~\ref{fig10} shows the experimental results of applying our method to \emph{Lena} image when block size ${ k }$=4. The original \emph{Lena} image ${ I }$ is presented in Fig.~\ref{fig-10-a}. After the content-owner completes the steps of reserving room and encryption, the encrypted image ${ I_{e} }$ in Fig.~\ref{fig-10-b} can be obtained. Then, the data-hider embeds additional data into the encrypted image ${ I_{e} }$ and generates a marked encrypted image ${ I_{ew} }$ in Fig.~\ref{fig-10-c}. It can be clearly seen that without the key, the content of the original image cannot be obtained by unauthorized users only from the encrypted image ${ I_{e} }$ or the marked encrypted image ${ I_{ew} }$, and both of images have security from the analysis in~\cite{li2019attacker}. Fig.~\ref{fig-10-d} displays the recovered \emph{Lena} image. The value of MSE (Mean square error) is 0, which proves that the recovered \emph{Lena} image is exactly the same as the original \emph{Lena} image. Since the size of embedding capacity is the key metric of RDHEI, the embedding rate (ER) is used as the evaluation index of embedding capacity. In addition, compared with other block sizes, ER is the highest when the block size ${ k }$ is 4, so our experiments are completed when ${ k }$ is 4. Table.~\ref{tb::tab1} shows the test results on three public datasets: BOSSbase~\cite{bas2011break}, BOWS-2~\cite{bas2017image} and UCID~\cite{schaefer2003ucid}. The average ERs are 3.498 ${ bpp }$ (bits per pixel), 3.393 ${ bpp }$ and 2.797 ${ bpp }$ respectively when block size ${ k }$=4. Moreover, the MSE of the recovered images on three datasets are 0, which indicates that each recovered image is exactly identical with the corresponding original image. In conclusion, RDHEI can be perfectly achieved by the proposed method.

\subsection{Security analysis}
\label{subsec::security analysis}
\par To prove the security of our method, we analyze it from two points on the basis of~\cite{dragoi2020security}.
\par
\begin{figure}[htbp]
  \centering
    \subfigure[The most significant bit plane]{
      \label{fig-11-a}
      \includegraphics[width= 1.65 in]{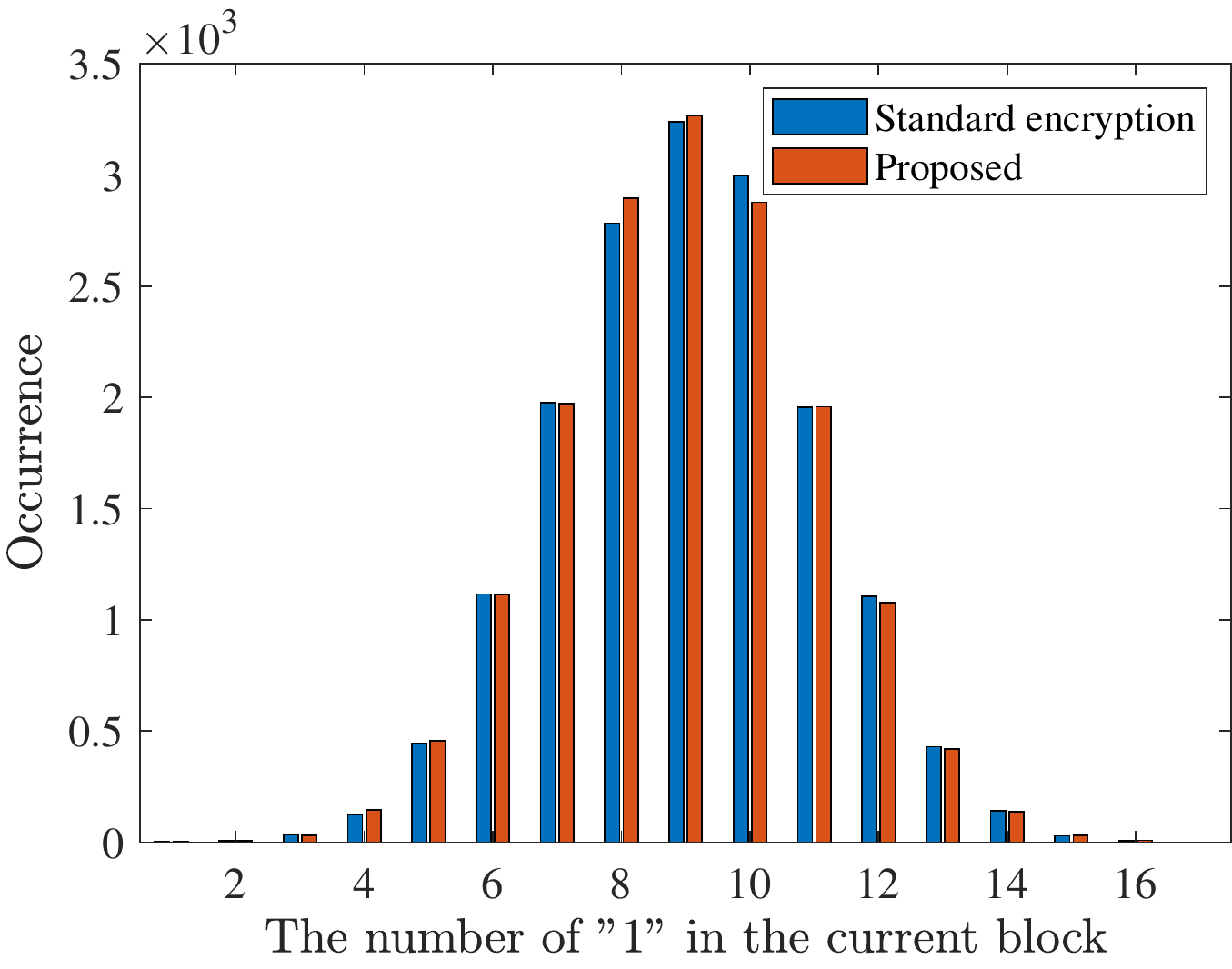}
    }
    \subfigure[The second significant bit plane]{
      \label{fig-11-b}
      \includegraphics[width= 1.65 in]{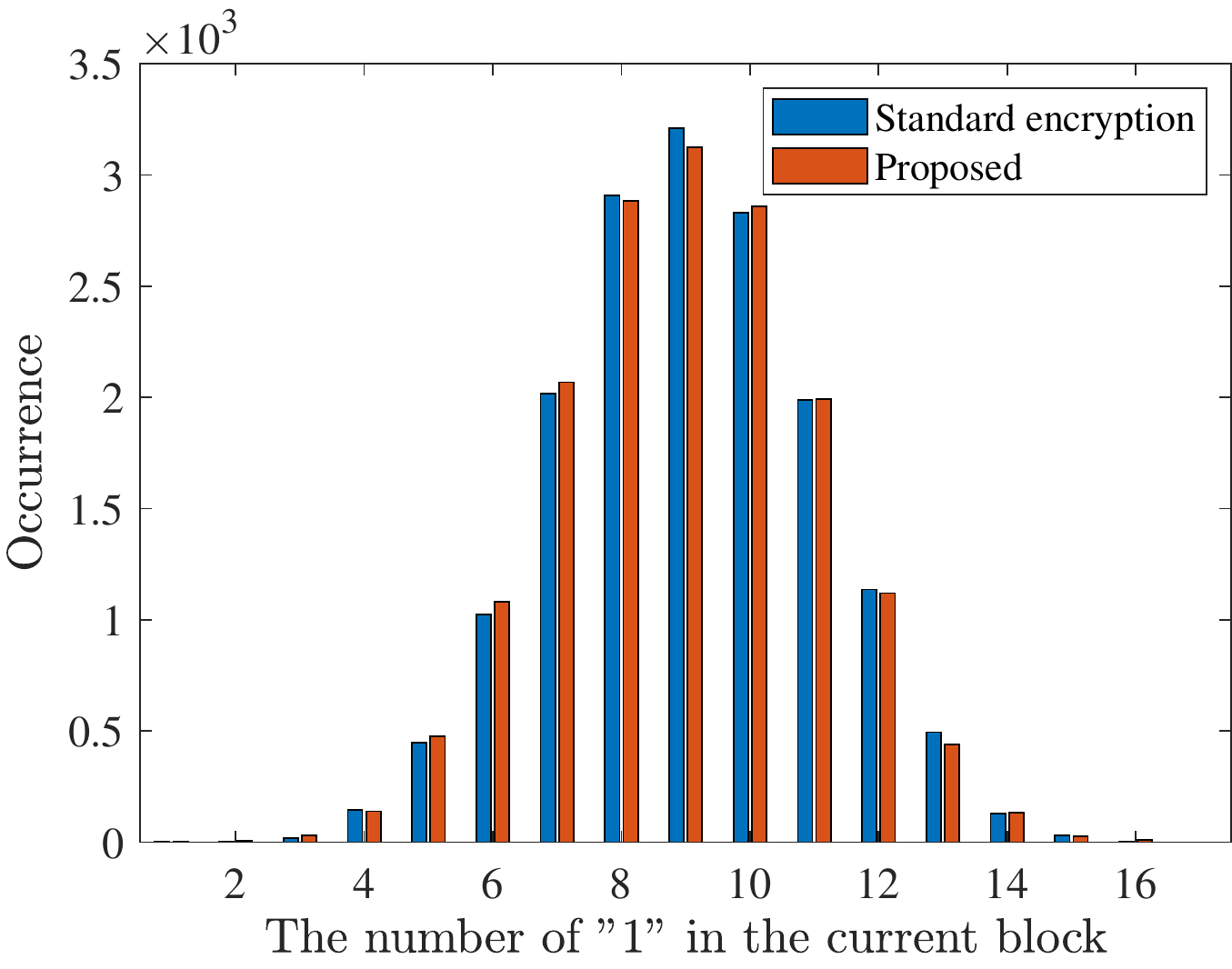}
    }
  \quad
    \subfigure[The third significant bit plane]{
      \label{fig-11-c}
      \includegraphics[width= 1.65 in]{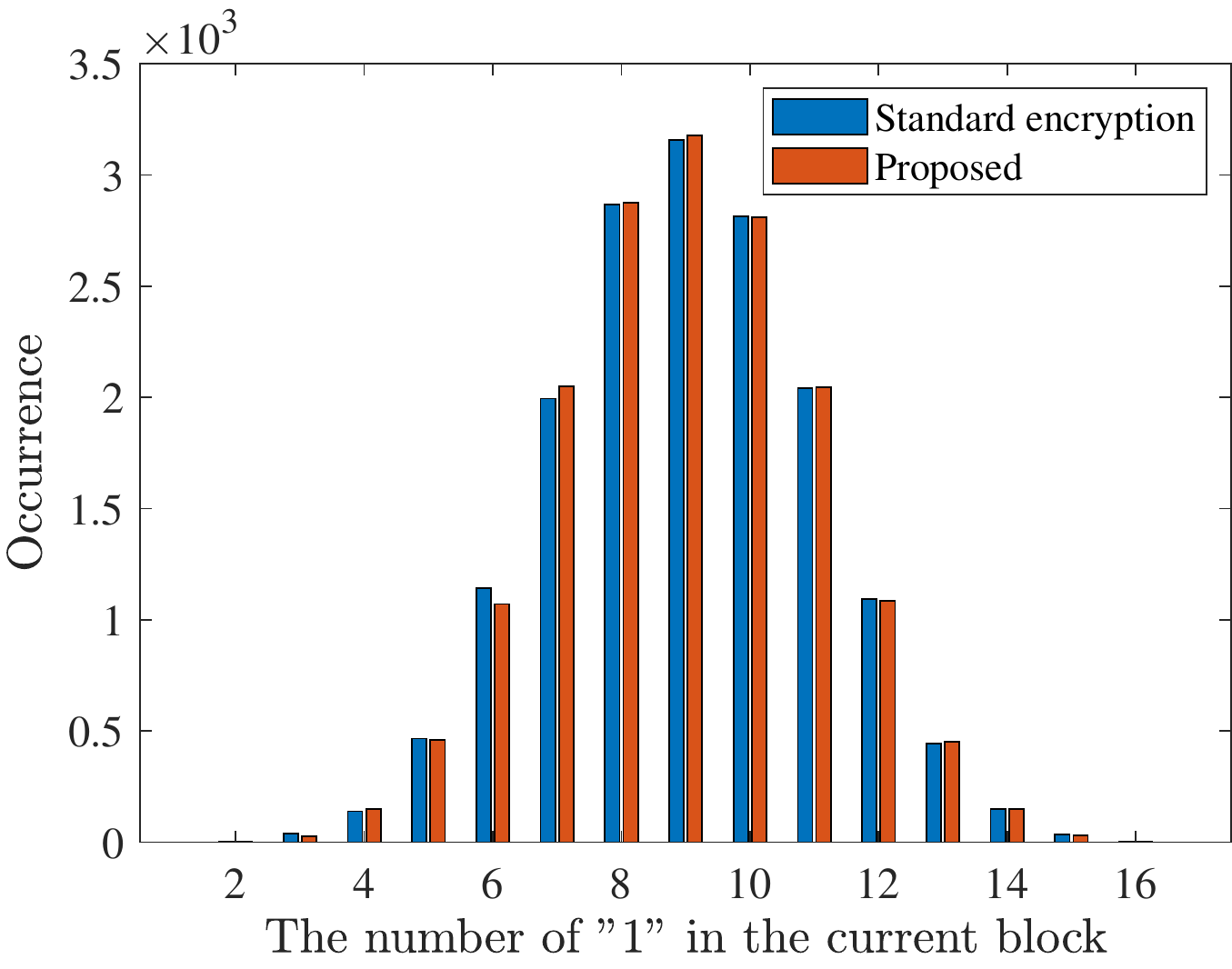}
    }
    \subfigure[The fourth significant bit plane]{
      \label{fig-11-d}
      \includegraphics[width= 1.65 in]{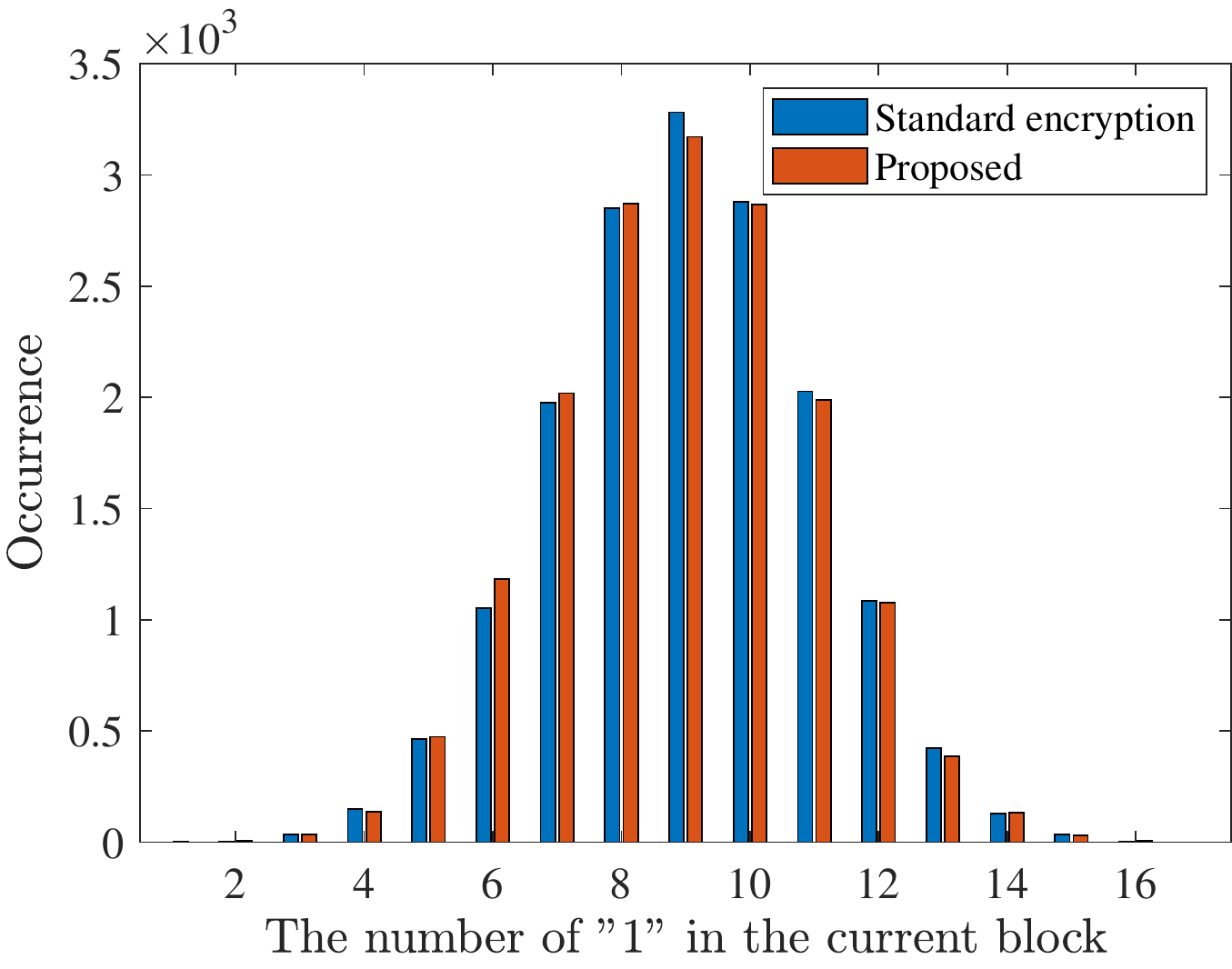}
    }
  \caption{ The number of occurrences of "1" in blocks between the standard encryption and the proposed method.}
  \label{fig11}
\end{figure}

 1) Embedding imperceptibility and data security. The report~\cite{dragoi2020security} pointed out that anyone who accesses the encrypted image can reveal the presence of auxiliary data by the distribution of runs of "1" in the encrypted MSB sequence~\cite{puteaux2018efficient}. Obviously, it is a serious security risk that unauthorized operators can easily obtain auxiliary data and get the location of the embedded data to alter it. Although the encrypted image with the auxiliary data needs to be transmitted to the data-hider in the proposed method too, there is no such risk. Since the auxiliary data is embedded into the block sized ${ 4 \times 4 }$ in the proposed method, an experiment has been designed to count the number of "1" in each ${ 4 \times 4 }$ block and the total number of blocks which corresponds to each number of "1". In Fig.~\ref{fig11}, four examples for several bit planes, the MSB plane, the second significant bit plane, the third significant bit plane and the fourth significant bit plane of \emph{Lena} image encrypted by the standard method and the proposed method are provided respectively (Standard encryption refers to the encryption of the original image without reserving room operation). It can be clearly seen that there is no significant difference in the number of occurrences of "1" in blocks between the standard encryption and the proposed method. This result is different from the method~\cite{puteaux2018efficient}, which proves that the security of our proposed method is close to that of the standard encryption. It is quite difficult to obtain the position of the embedded data by analyzing the characteristics of encrypted images, hence embedding invisibility and data security can be achieved in our method.

\begin{figure}[htbp]
  \centering
    \subfigure[]{
      \label{fig-12-a}
      \includegraphics[width= 1.65 in]{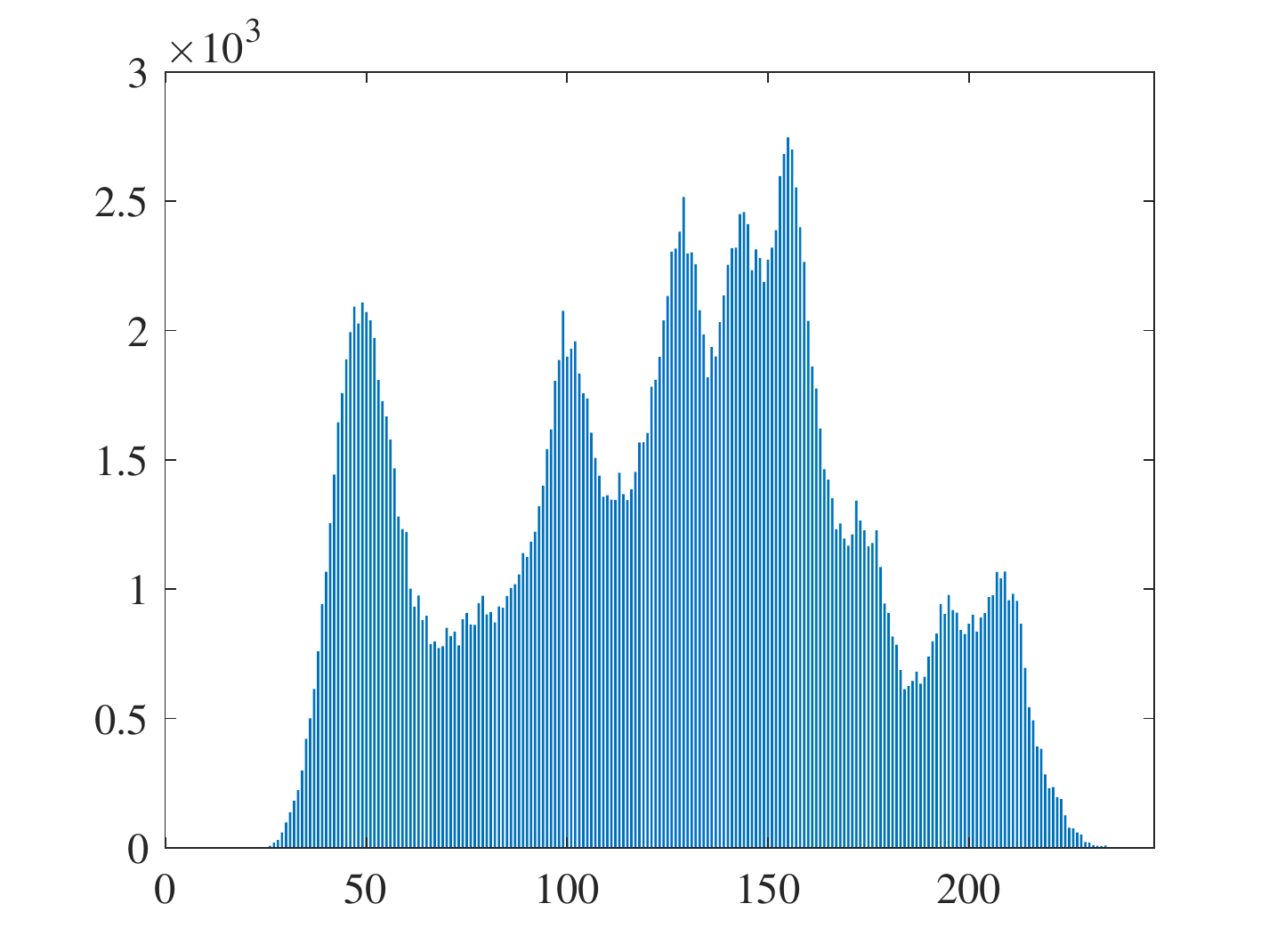}
    }
    \quad
    \subfigure[]{
      \label{fig-12-b}
      \includegraphics[width= 1.65 in]{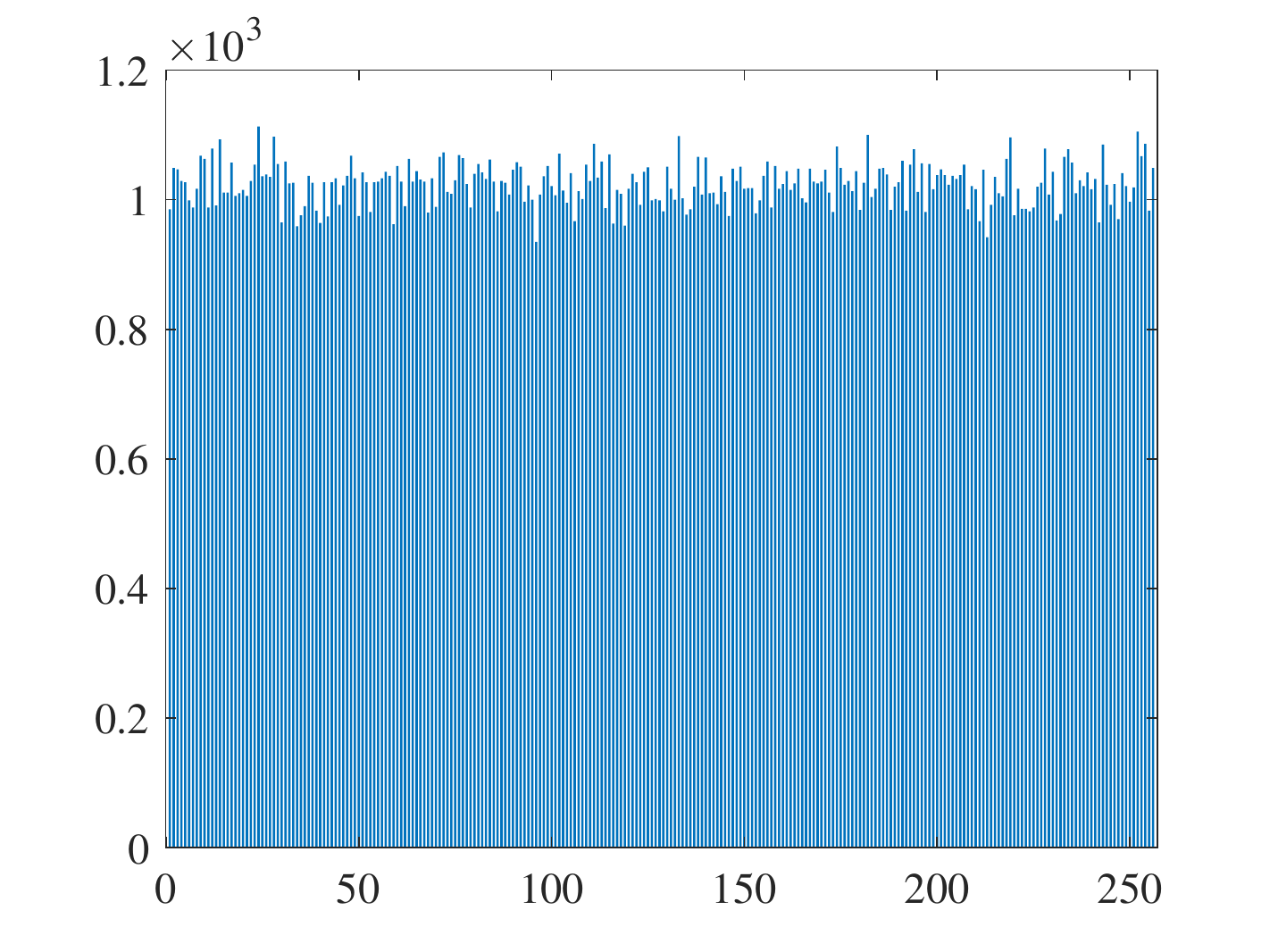}
    }
    \subfigure[]{
      \label{fig-12-c}
      \includegraphics[width= 1.65 in]{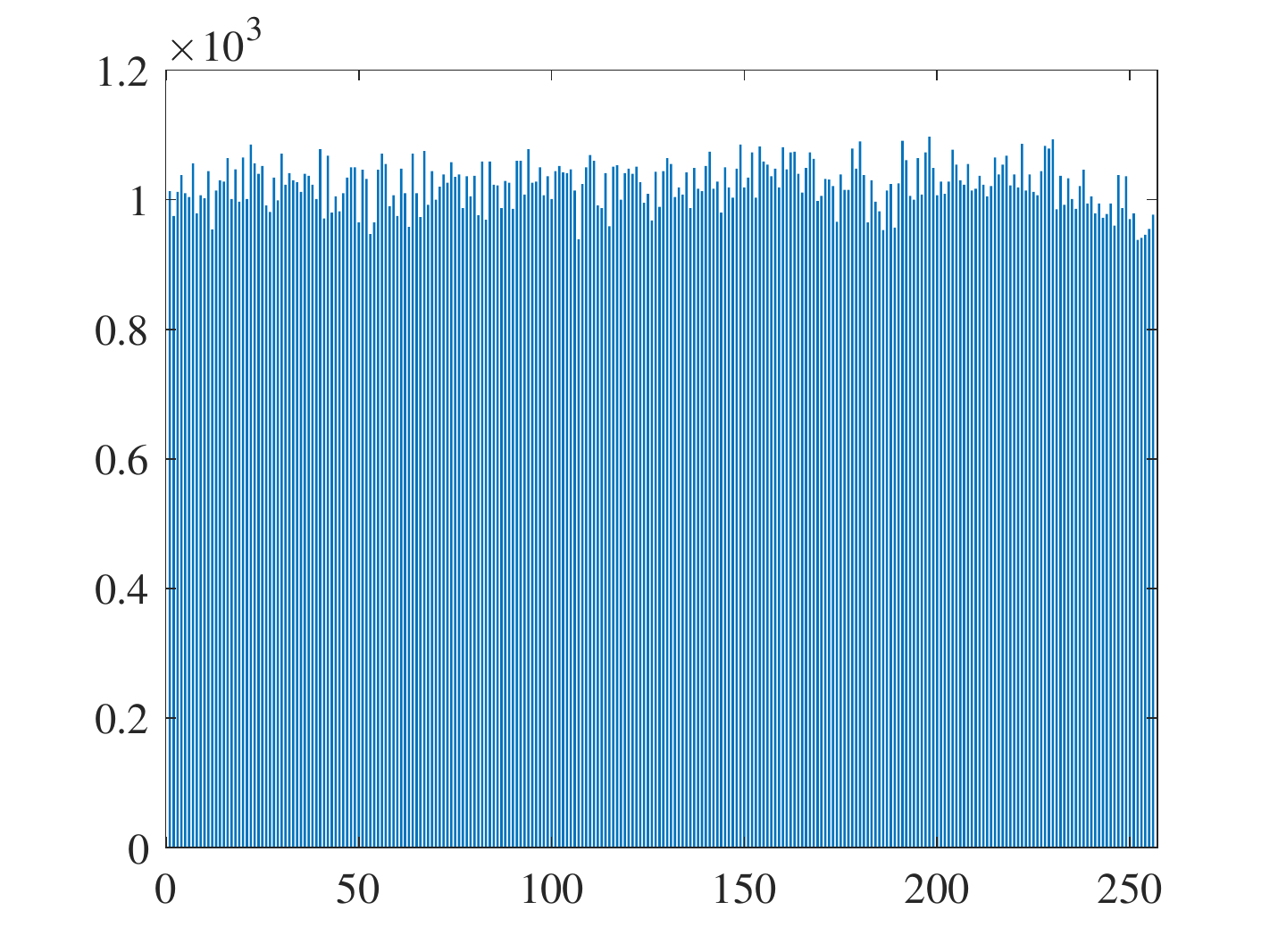}
    }
  \caption{ Histograms of \emph{Lena} image: (a) the original image; (b) the standard encrypted image; (c) the proposed encrypted image.}
  \label{fig12}
\end{figure}

\par 2) Unauthorized access to image content. The report~\cite{dragoi2020security} pointed out that when an unauthorized operator or data-hider receives the encrypted image, the original image content can be obtained through the extracted auxiliary data. Since neither data-hiders nor unauthorized operators have the right to access the original image, this is a security risk. It should be emphasized that the location of embedded data is only a part of auxiliary data in the proposed method. If the data-hider wants to obtain the original image content without authorization, he can only get all the auxiliary data by analyzing the characteristics of the encrypted image. As shown in Fig.~\ref{fig-12-a}, the histogram of the original image contains obvious feature information. It can be seen in Fig.~\ref{fig-12-b}, after standard encryption (Standard encryption refers to the encryption of the original image without auxiliary data), the distribution of pixels becomes uniform. Fig.~\ref{fig-12-c} shows the encrypted image in the proposed method, and the distribution of pixels becomes uniform too. The statistical features of the two encrypted images are both significantly different from the original image. Actually, for each grayscale image with a size of ${ M \times N }$, we have used an ${ 8 \times M \times N }$ binary sequence to encrypt it. Without the correct encryption key, the probability of correctly obtaining the binary sequence to decrypt the encrypted image is ${ \frac{1}{2^{8 \times M \times N}}}$, which is almost impossible. Therefore, the proposed method has strongly security, in which the original image content is well protected and it is almost impossible for the operators without encryption key to obtain it.

\subsection{Performance analysis}
\label{subsec::performance analysis}
\par In this section, we analyze the performance of proposed method according to the contributions mentioned earlier. The first is about reversibility. It can be seen from Table.~\ref{tb::tab1} that after using the proposed method, the MSE of the restored image on the three datasets is all 0. It means that our method is fully reversible. Next is the analysis on embedding capacity, which is the key indicator of RDHEI. As talked in the previous sections, the pixel prediction scheme of binary images can be directly applied to the bit plane of grayscale images, but experimental results demonstrate the embedding capacity is unsatisfactory. However, our method can achieve high embedding capacity.
\par One of the reasons is that our method doesn’t generate too much auxiliary data to occupy much embeddable room, and the auxiliary data can be perfectly compressed. As mentioned in Section~\ref{subsec::room reservation}, the auxiliary data in this paper is mainly composed of ${ L_1 }$ and ${ L_2 }$. ${ L_1 }$ is the label map of overflow pixels and total pixels, and the original size is the same as the original image. However, the percentage of overflow pixels and total pixels is extremely low. For example, in the five commonly used images, the percentages are 0.04$\%$, 1.42$\%$, 0.16$\%$, 0.09$\%$ and 0.02$\%$ respectively. Thus, we set tag 1 for overflow pixels, 0 for other pixels, and ${ L_1 }$ with a large number of 0 and a few 1 is generated. It can be perfectly compressed by arithmetic coding because the sequence ${ L_1 }$ is sparse. ${ L_2 }$ is another label map, which shows the original positions of UBs and NUBs before the bit plane is rearranged. Since there are more UBs and less NUBs in most PE bit planes, we also set tag 1 for NUBs, 0 for UBs to generate ${ L_2 }$. In the same way, arithmetic coding can be used to compress ${ L_2 }$. After compression, there is more room in bit planes to embed additional data, which further improves the ER of our method.
\par
\begin{figure}[htbp]
  \centering
    \subfigure[]{
      \label{fig-13-a}
      \includegraphics[width= 1.65 in]{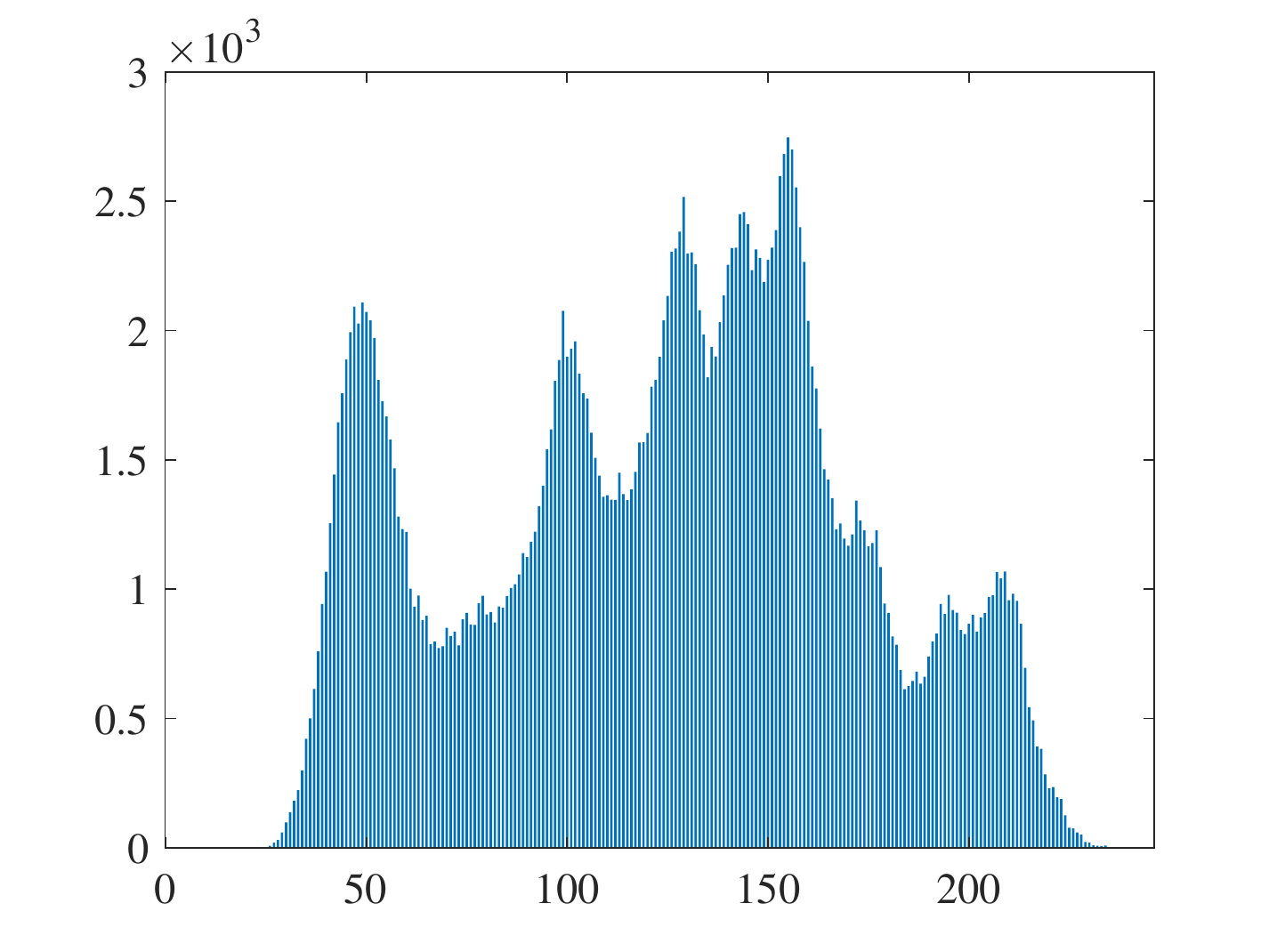}
    }
    \subfigure[]{
      \label{fig-13-b}
      \includegraphics[width= 1.65 in]{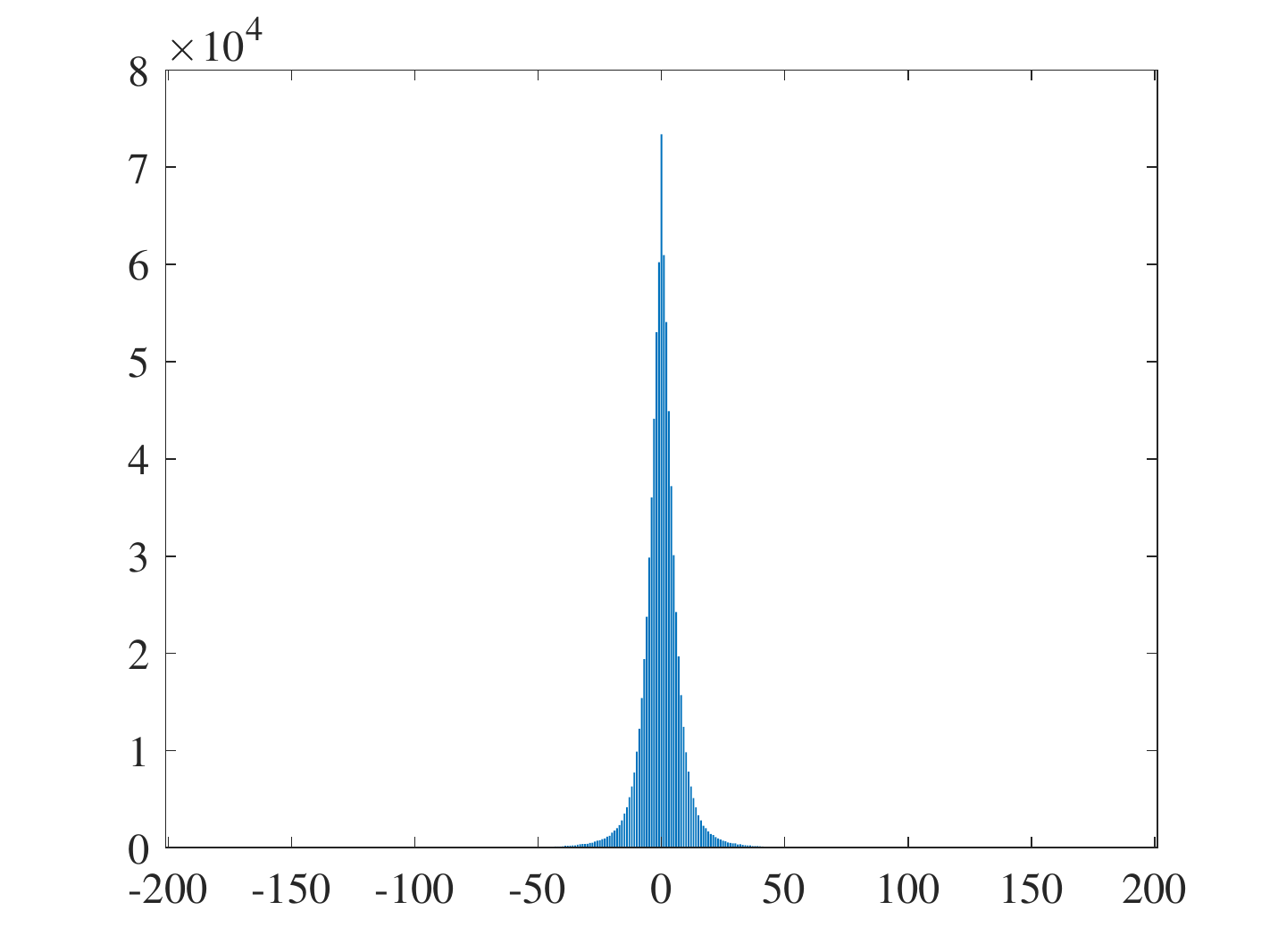}
    }
  \caption{ Histograms of \emph{Lena} image: (a) the original image; (b) the prediction error image.}
  \label{fig13}
\end{figure}

 Another reason is that we reserve the room of PE bit planes instead of the original bit planes. In the proposed method, we use the MED predictor to calculate predicted values. After the calculation, as shown in Fig.~\ref{fig13}, the distribution of PEs is more concentrated than that of original pixel values. In other words, it makes more values in bit planes being identical and means that more data can be embedded into UBs. In summary, the proposed method exploits spatial redundancy better and makes the available embedded room larger, hence the performance of it is more satisfactory.


\begin{figure*}[!ht]
  \centering
  \includegraphics[width=1\textwidth]{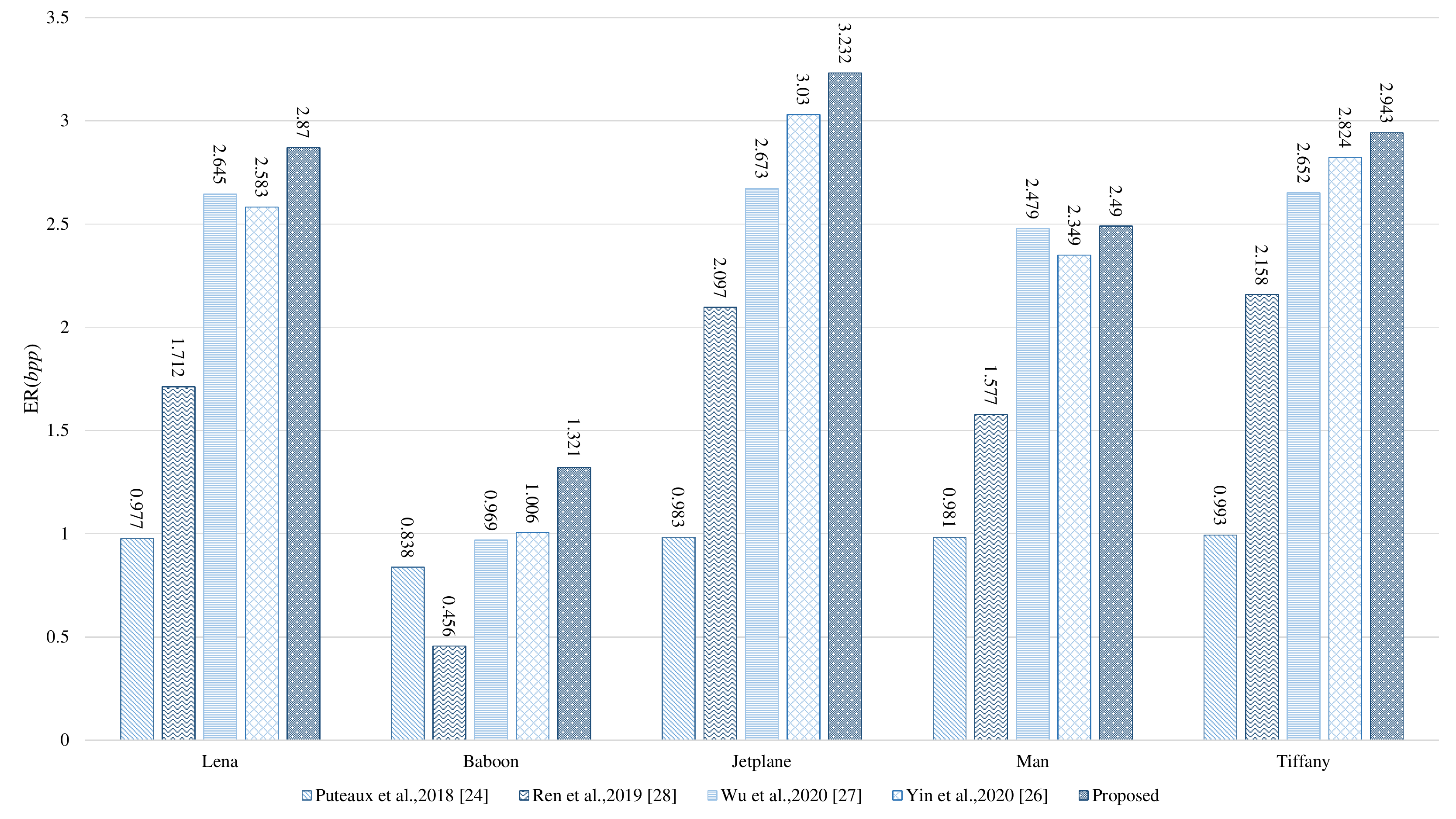}
  \caption{Comparison of ER ${ (bpp) }$ on five test images.}
  \label{fig14}
\end{figure*}
\begin{figure*}[!ht]
  \centering
  \includegraphics[width=1\textwidth]{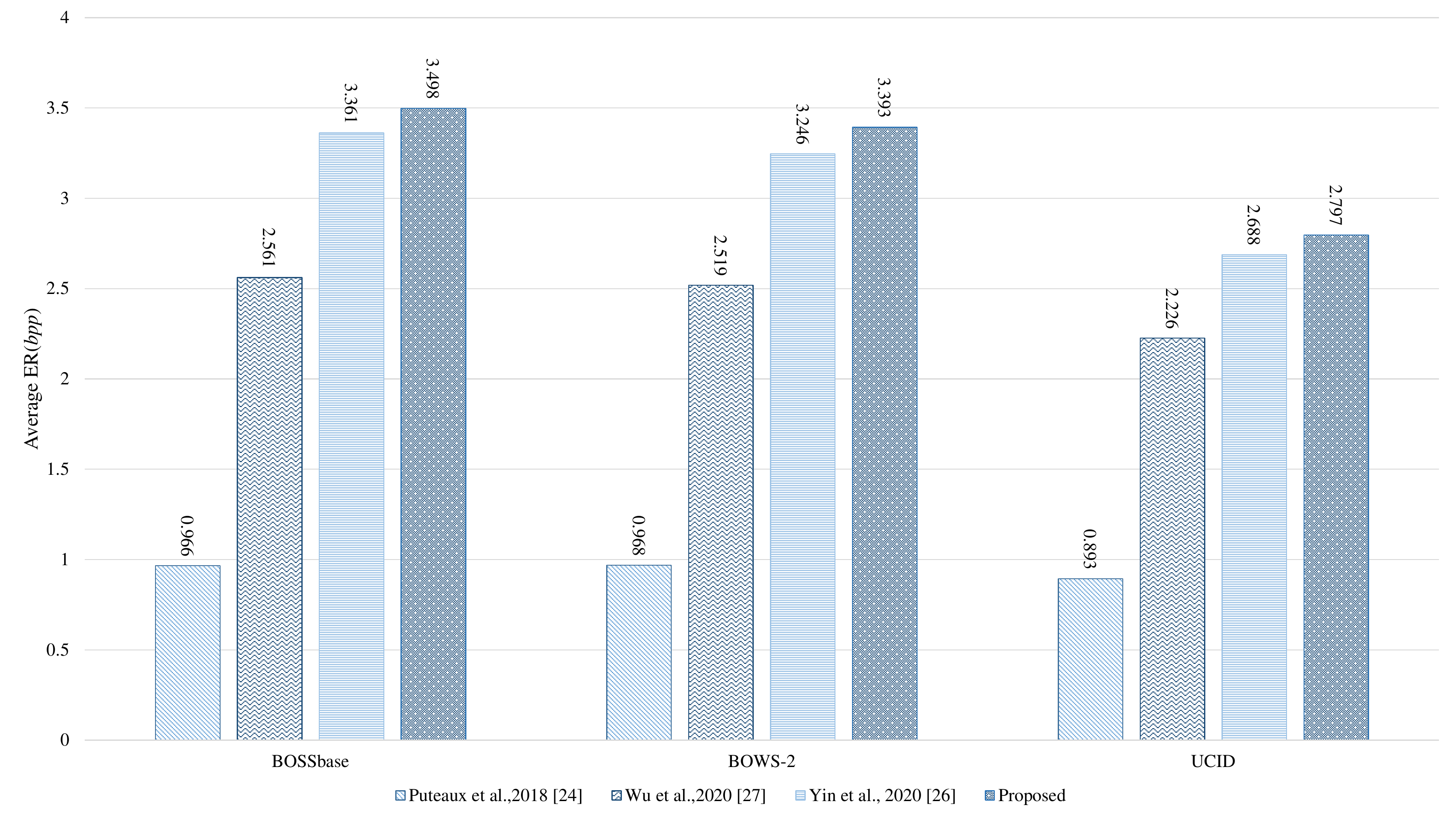}
  \caption{Average ER ${ (bpp) }$ comparison in different datasets: BOSSBase ~\cite{bas2011break}, BOWS-2~\cite{bas2017image}, and UCID~\cite{schaefer2003ucid}.}
  \label{fig15}
\end{figure*}

\begin{table}[]
  \centering
  \caption{Experimental results on three image databases.}
  \label{tb::tab1}
  \begin{tabular}{@{}ccc@{}}
    \toprule
    \textbf{Database} & \multicolumn{1}{l}{\textbf{MSE}} & \textbf{Average ER( $ bpp $)} \\ \midrule
    BOSSbase          & 0                                & 3.498                         \\
    BOWS-2            & 0                                & 3.393                         \\
    UCID              & 0                                & 2.797                         \\ \bottomrule
  \end{tabular}
\end{table}

\subsection{Comparison with state-of-the-art methods}
\label{subsec::Comparison}
\par To better verify the good performance of the proposed method, this section compares ER  with three latest state-of-the-art RDHEI methods~\cite{puteaux2018efficient,yin2019reversible,wu2019improved}. As shown in Fig.~\ref{fig14}, five commonly used images are tested by using these methods, and it can be seen that our method outperforms the other three methods with highest ER.
\par Different from our method of embedding data into multiple MSBs, Puteaux et al.~\cite{puteaux2018efficient} only embedded data into the MSB, and other bits are not fully utilized. Hence the ER of their method is very low. For example, in~\cite{puteaux2018efficient}, the ERs of five commonly used images are only 0.977 ${ bpp }$, 0.838 ${ bpp }$, 0.983 ${ bpp }$,0.981 ${ bpp }$ and 0.993 ${ bpp }$. In~\cite{wu2019improved}, Wu et al. suggested marking pixels with a parameter binary tree and dividing them into two categories, which would make some pixels unable to embed additional data and affect the performance of embedding capacity. The ERs of five test images in their paper are 2.645 ${ bpp }$, 0.969 ${ bpp }$, 2.673 ${ bpp }$, 2.479 ${ bpp }$ and 2.652 ${ bpp }$. In addition, Yin et al.~\cite{yin2019reversible} uses multiple MSBs, but the auxiliary data generated by this method is very large and cannot be compressed well, so the embeddable room of each bit plane is still not well utilized. When using the method of Yin et al.~\cite{yin2019reversible}, the ERs of five test images are 2.583 ${ bpp }$, 1.006 ${ bpp }$, 3.03 ${ bpp }$, 2.349 ${ bpp }$ and 2.824 ${ bpp }$. However, the experimental results of five images in this paper are 2.87 ${ bpp }$, 1.321 ${ bpp }$, 3.232 ${ bpp }$, 2.49 ${ bpp }$, 2.943 ${ bpp }$ respectively, which are higher than previous methods. Finally, to better prove the performance of our method, we have directly applied the binary image method of Ren et al.~\cite{ren2019reversible} to grayscale images. This method performs reserving operations directly on original bit planes. Since there are only a few embeddable positions in original bit planes compared to the prediction error bit planes proposed in our paper, the embedding capacity is unsatisfactory. Furthermore, this method will generate too much auxiliary data, which is not sparse sufficiently. Therefore, the auxiliary data cannot be well compressed and direct application will not achieve high embedding capacity. In other word, direct application can only achieve high embedding capacity without embedding auxiliary data. However, it is impossible to realize reversibility without embedding auxiliary data. On the contrary, if we want to achieve reversibility, the embedded capacity cannot be satisfactory. Our experimental results show that, as mentioned in the previous contributions, we have solved the problem that high embedding capacity and reversibility cannot coexist. Taking the \emph{Lena} image as an example, as shown in Fig.~\ref{fig14}, the ER is only 1.712 ${ bpp }$. It is worth mentioning the ER of \emph{Lena} image when using our method is 2.87 ${ bpp }$. At last, we also compared the ER with these state-of-the-art methods on three datasets: BOSSbase~\cite{bas2011break}, BOWS-2~\cite{bas2017image} and UCID~\cite{schaefer2003ucid}. As shown in Fig.~\ref{fig15}, it is obvious that the proposed method has a higher ER on these datasets than other methods.

\section{Conclusions}
\label{sec::Conclusion}
\par This paper proposes a new method of reversible data hiding in encrypted images based on pixel prediction scheme and multi-MSB planes rearrangement, which achieves a higher embedding capacity compared with previous methods. It is worth noting that compared with previous methods, we can make better use of the correlation of adjacent pixels. The experimental results show that the proposed method not only has a significantly increase in net payload, but also has reversibility. Moreover, the proposed method achieves higher ER than using pixel prediction scheme directly on grayscale images. Main reasons for better performance are: First, the distribution of prediction errors is more concentrated than that of original pixel values, and more embeddable positions of the bit plane can be used to embed data. Next, we make full use of the correlation of adjacent pixels and compress auxiliary data with sparse features via arithmetic coding, which makes more embeddable positions in the bit planes.
\par In the future work, we will continue to work on improving the performance of our algorithm. For example, the number of auxiliary data can be reduced as much as possible, and a more accurate prediction scheme can be used to improve the embedding capacity. In addition, we hope that this paper will enable readers to understand the possibility of applying binary image algorithm to grayscale images, and the idea of improving performance during application. It is very practical to explore more general algorithms based on binary images and grayscale images in the future.

\section*{Acknowledgments}
\label{sec::Acknowledgments}
\par This research work is partly supported by National Natural Science Foundation of China (61872003, U20B2068, 61860206004).

\bibliography{elsarticle-template}

\end{document}